\documentclass[reprint,prb]{revtex4-2}

\usepackage{comment}
\usepackage{xcolor}
\usepackage{graphicx}
\usepackage{amsmath}
\usepackage{tabularx}

\newcommand{\vect}[1]{\boldsymbol{#1}}

\begin{document}

\title{Proton Quantum Effects in H$_3$S Electronic Structure: A Multicomponent DFT study via Nuclear-Electronic Orbital Method }

\author{Jianhang Xu}
\author{Aaron M. Schankler}
\affiliation{Department of Chemistry, University of North Carolina at Chapel Hill, Chapel Hill, North Carolina 27599, USA}
\author{Yosuke Kanai}
\email{ykanai@unc.edu}
\affiliation{Department of Chemistry, University of North Carolina at Chapel Hill, Chapel Hill, North Carolina 27599, USA}
\affiliation{Department of Physics and Astronomy, University of North Carolina at Chapel Hill, Chapel Hill, North Carolina 27599, USA}

\date{\today}

\begin{abstract}
We investigate the impact of the quantum effects of protons on the electronic structure of high-pressure H$_3$S, a benchmark hydrogen-rich superconductor with a critical temperature ($T_c$) exceeding 200 K.
Using Nuclear-Electronic Orbital Density Functional Theory (NEO-DFT), we treat hydrogen nuclei quantum mechanically on the same footing as electrons within a first-principles framework.
Our calculations reveal that nuclear quantum effects (NQEs) induce subtle modifications to the electronic band structure and density of states (DOS) near the Fermi energy, including features associated with van Hove singularities.
However, the resulting changes in the DOS would increase $T_c$ by only a few percent.
{On the other hand, calculations of the phonon dispersion with the NEO-DFT method show large changes in the hydrogen-dominated phonons that arise from a stiffening of the S-H bonds due to NQEs. 
These findings imply that the experimentally observed reduction in $T_c$ upon deuteration arises predominantly from changes in the phonon properties while NQEs-induced modifications to the electronic structure itself are minimal.}
\end{abstract}

\maketitle


\section{Introduction}

\par 
First-principles theory for accurately predicting the superconducting properties of real materials has advanced substantially in recent years, particularly for conventional phonon-mediated superconductors \cite{calandra2007anharmonic,errea2016approaching,giustino2017electron,zurek2019high,flores2020perspective}. 
The foundational Bardeen-Cooper-Schrieffer (BCS) theory highlights the importance of electron-phonon interaction and the electronic density of states (DOS) at the Fermi level in enabling Cooper pairing for superconductivity \cite{bardeen1957theory}.
Building upon this framework, Migdal-Eliashberg theory provides a quantitative description of superconductivity such that the critical temperature can be computed with key parameters from density functional theory (DFT) and density functional perturbation theory (DFPT).
Ongoing efforts continue to refine this theoretical framework by incorporating strong-coupling corrections \cite{lee2023electron}, quantum anharmonic phonon effects \cite{mishra2025electron}, and improved descriptions of the Coulomb interaction \cite{pellegrini2024ab}, etc.
\par 
Sulfur hydride (H$_3$S) 
is a paradigmatic example of how first-principles theory provided a quantitative prediction of superconductivity.
In 2004, \citeauthor{ashcroft2004hydrogen} argued that hydrogen-dominant materials could be suitable candidates for high-$T_c$ phonon-mediated superconductors due to the light mass of hydrogen and the resulting strong electron-phonon coupling, particularly under high pressure \cite{ashcroft2004hydrogen}.
A decade later, \citeauthor{duan2014pressure} applied an evolutionary structure search combined with DFT calculations to identify the \textit{Im$\bar{3}$m} phase of H$_3$S as a stable high-temperature superconductor under high pressure \cite{duan2014pressure}.
Using the Allen-Dynes equation, a modified McMillan equation, they estimated a superconducting critical temperature ($T_c$) in the range of 191 K to 204 K at around 200 GPa.
This theoretical prediction was soon confirmed experimentally by \citeauthor{drozdov2015conventional} in 2015, who reported superconductivity at 203 K under similar conditions \cite{drozdov2015conventional}.
Recent experimental confirmations include even the direct measurement of the temperature-dependent superconducting gap \cite{du2025superconducting}. 
As expected for H$_3$S where phonons are responsible for pairing of electrons as Cooper pairs, the critical temperature is lowered significantly when deuterated (i.e., D$_3$S) \cite{errea2015high}. 
The semi-empirical McMillan formula directly relates $T_c$ to 
the electron-phonon (el-ph) coupling parameter $\lambda$ and the logarithmic average of the phonon frequencies $\omega_\text{log}$, which are both obtained from Eliashberg spectral function, $\alpha^2 F$, and the phonon frequencies. 
In recent years, Migdal-Eliashberg (ME) theory based on the Green's function formalism uses $\alpha^2 F$ to compute the superconducting gap to determine $T_c$ from first principles \cite{carbotte1990properties,giustino2017electron}. 
According to these theoretical formulations, a significant reduction in $T_c$ is expected upon deuteration. 
H$_3$S has served as a benchmark for examining and refining first-principles computational methods for predicting superconductivity \cite{errea2015high,bernstein2015superconducts,papaconstantopoulos2015cubic,quan2016van,ortenzi2016band,durajski2017first,szczkesniak2018unusual}.

\par
With the small mass of hydrogen atoms, nuclear quantum effects (NQEs) represent another key feature of hydrogen-rich materials like H$_3$S.
The zero-point energy (ZPE) has been shown to significantly impact the phase and structural stability of H$_3$S and to be responsible for the unique symmetry of the S-H-S bonds \cite{errea2015high,jarlborg2016breakdown,errea2016quantum,sano2016effect,taureau2024quantum}.
In our earlier work \cite{xu2022nuclear}, the proton NQEs were found to change the electronic structure details of hydrogen boride sheets, which also possess the same unique H-bridging bonds (B-H-B); the band structure and the density of states (DOS) are noticeably altered. 
We also note that deuteration is known to shift and reshape the DOS in organic solids experimentally \cite{cai2024deuteration}.
{In addition to the mass-related renormalization of properties like electron-phonon coupling \cite{akashi2015first}, $T_c$ change from deuteration could also result partly from the electronic structure changes due to NQEs.}
The Eliashberg spectral function is proportional to the electronic DOS at the Fermi level, $N(\epsilon_F)$, and the well-known Allen-Dynes limit of the strong coupling directly relates the transition temperature to the DOS via $ T_c \propto \sqrt{N(\epsilon_F)}$ \cite{allen1975transition}.
The effect of proton quantization in first-principles calculations is particularly relevant for H$_3$S because its high $T_c$ has been attributed in part to the presence of van Hove singularities (vHS) near the Fermi energy in the DOS \cite{quan2016van}.
These singularities enhance the DOS and contribute to stronger electron-phonon coupling near the Fermi level, but they are also sensitive to details of the electronic structure.
Whether and to what extent NQEs, including the correlation between electrons and quantum-mechanical protons, alter these key electronic structure features is an open question. 

{Previous studies have examined NQEs in hydride superconductors primarily through perturbative zero-point renormalization of the electronic structure within the Born-Oppenheimer framework \cite{sano2016effect}.
To our knowledge, a fully self-consistent, multicomponent first-principles treatment in which quantum protons explicitly enter the ground-state energy functional, has not been applied to address this lingering question from electronic structure theory.} 
The multicomponent DFT \cite{capitani1982non,kreibich2001multicomponent} provides an ideal theoretical formalism, allowing us to transcend the usual Born-Oppenheimer approximation in describing the quantum-mechanical correlation between electrons and protons.
The nuclear-electronic orbital (NEO) method \cite{webb_multiconfigurational_2002,hammes-schiffer_nuclearelectronic_2021} provides a particularly convenient approach for implementing the multicomponent DFT within the Kohn-Sham (KS) ansatz \cite{pak_density_2007,chakraborty_development_2008}.
In the NEO-DFT approach, selected nuclei are treated quantum mechanically on the same footing as electrons,
and their wavefunctions are solved self-consistently within the multicomponent KS formalism \cite{capitani1982non,kreibich2001multicomponent}.
The correlation between electrons and quantum protons is taken into account via the electron-proton correlation functional \cite{udagawa2006h,chakraborty_development_2008}, which goes beyond the classical electrostatic interaction.
The NEO method naturally takes into account the nuclear quantum effects, including delocalization and zero-point energy, while retaining a similar computational cost to standard DFT \cite{pavo2020multicomponent}.
It has been successfully used for molecular systems to study various properties such as proton affinity and proton-coupled electron transfer \cite{brorsen2017multicomponent}.
More recently, it has also been adapted for studying extended periodic condensed-phase systems \cite{xu2022nuclear}.

\par In this work, we apply our newly developed NEO-DFT method to study the electronic structure of high-pressure H$_3$S in the \textit{Im$\bar{3}$m} phase
using both hydrogen (H) and deuterium (D) nuclei.
The remainder of this paper is organized as follows.
In Sec. \ref{sec:method}, we briefly introduce the periodic NEO-DFT method and describe the computational details.
In Sec. \ref{sec:result}, we present and analyze nuclear quantum effects and isotope effects on the electronic band structure, the density of states (DOS), and the Fermi surface.
We also discuss the role of electron-proton correlation in NEO-DFT and examine the localization behavior of H and D nuclei, as well as the proton quantization effects on phonons, before concluding in Sec. \ref{sec:conclusion}.

\section{Theoretical Method}
\label{sec:method}
\subsection{NEO-DFT Method}

\begin{figure}[t]
    \centering
    \includegraphics[width=0.3\textwidth]{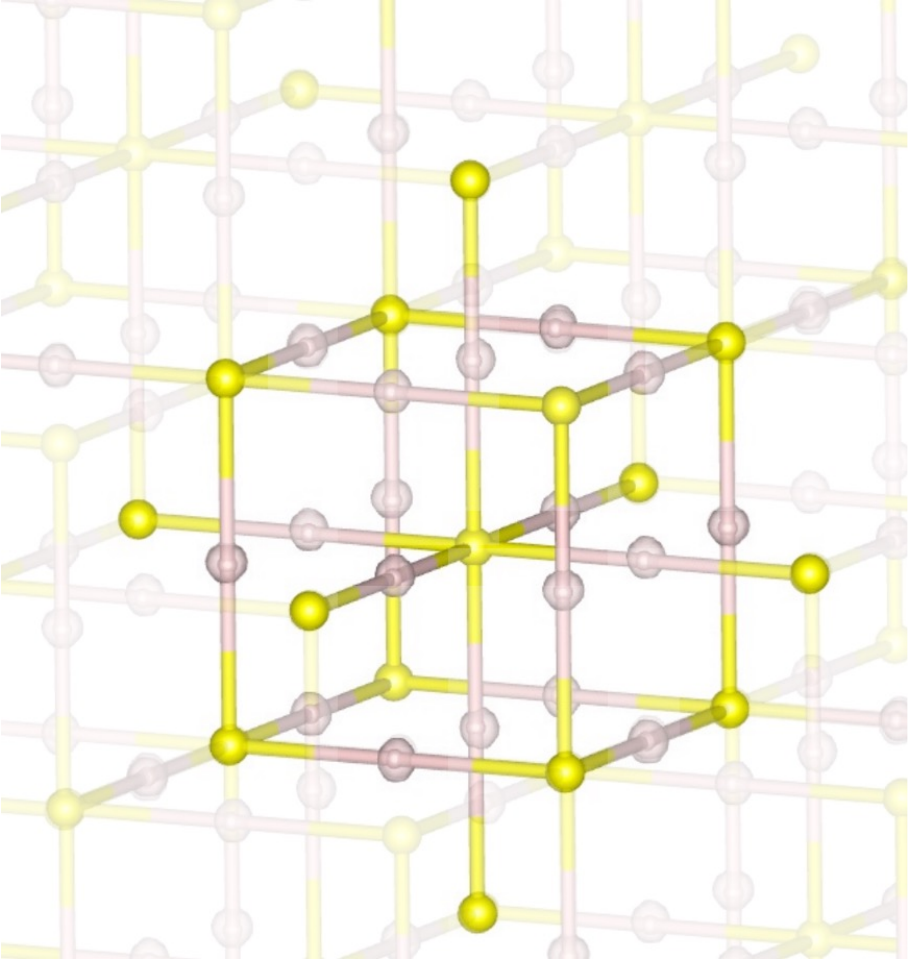}
    \caption{Crystal structure of H$_3$S in \textit{Im$\bar{3}$m} space group, having the conventional face-centered cubic unit cell. The large yellow spheres represent classical sulfur atoms, and the pink isosurfaces depict the quantum proton density obtained from NEO-DFT calculations.}
    \label{fig:Structure}
\end{figure}

\par 
As discussed in Ref. \citenum{xu2022nuclear}, the NEO approach enables us to formulate multicomponent DFT via a set of coupled electron-nucleus KS equations. 
We select hydrogen atomic nuclei in H$_3$S as quantum particles, and they are modeled quantum-mechanically on an equal footing as electrons.
For brevity, let us refer to the H atom nuclei as protons, even though deuteration is also considered in this work.
The coupled KS equations are
\begin{equation}
\label{eq:KS}
\begin{aligned}
    \hat{H}^e_{\bf{k}} \psi^e_{i,\bf{k}} (\mathbf{r}^e) &= \left[-\frac{1}{2}\nabla^2+U_{\text{eff}}^e(\mathbf{r}^e)\right] \psi^e_{i,\bf{k}} = \epsilon^e_{i,\bf{k}} \psi^e_{i,\bf{k}} (\vect r^e), \\
    \hat{H}^p \psi^p_j (\mathbf{r}^p)&= \left[-\frac{1}{2M^p}\nabla^2+U_{\text{eff}}^p(\mathbf{r}^p)\right] \psi^p_j = \epsilon^p_j \psi^p_j (\vect r^p), 
\end{aligned}
\end{equation}
where $M^p$ is the proton mass, $\psi^e_{i,\bf{k}}$, $\psi^p_{j}$ and $\epsilon^e_{i,\bf{k}}$, $\epsilon^p_{j}$ are the KS orbitals and eigenvalues, respectively.
The superscript $e,p$ denotes the electronic and protonic subsystems.
The effective potentials $U_{\text{eff}}$ are defined as
\begin{equation}
    \begin{aligned}
    U_{\text{eff}}^e(\mathbf{r}^e) =& -V_{\text{ext}}(\mathbf{r}^e) - V_{\text{es}}^{e}(\mathbf{r}^e)-V_{\text{es}}^{p}(\mathbf{r}^e)\\
    &+\frac{\delta E_{\text{xc}}^{e}[\rho ^e]}{\delta \rho^e}(\mathbf{r}^e)+\frac{\delta E_{\text{epc}}[\rho^e,\rho^p]}{\delta \rho^e}(\mathbf{r}^e), \\
    U_{\text{eff}}^p(\mathbf{r}^p) =& V_{\text{ext}}(\mathbf{r}^p) + V_{\text{es}}^{e}(\mathbf{r}^p)+V_{\text{es}}^{p}(\mathbf{r}^p)\\
    &+\frac{\delta E_{\text{xc}}^{p}[\rho ^p]}{\delta \rho^p}(\mathbf{r}^p)+\frac{\delta E_{\text{epc}}[\rho^e,\rho^p]}{\delta \rho^p}(\mathbf{r}^p),
\end{aligned}
\end{equation}
where $V_{\text{ext}}$, $V_{\text{es}}^{e}$, and  $V_{\text{es}}^{p}$ denote the electrostatic potentials of the classical nuclei, electrons, and quantum protons, respectively. 
$E_{\text{xc}}^e$ and $E_{\text{xc}}^p$ are the exchange-correlation (XC) energy functionals for the electronic and the protonic subsystems. 
$E_{\text{epc}}$ represents the electron-proton correlation energy,
which is given as a functional of both the electronic and protonic densities in multicomponent DFT \cite{brorsen2017multicomponent,yang2017development,brorsen2018alternative,tao2019multicomponent}. 
Generally speaking, the KS wave functions for quantum protons are highly localized in real space such that Brillouin zone (BZ) integration can be safely neglected for the protonic subsystem \cite{xu2022nuclear}.
For D$_3$S calculations, the proton mass is replaced with that of the deuterium in Eq. \ref{eq:KS}.


\subsection{Computational Details}
\par 
We used the experimental crystal structure of H$_3$S in the \textit{Im$\bar{3}$m} phase with a lattice constant of 2.98504 \AA, corresponding to a pressure of 200 GPa (see Fig. \ref{fig:Structure}) \cite{drozdov2015conventional}, {for all standard DFT and NEO-DFT calculations of H$_3$S and its isotopologue D$_3$S}.
All simulations were carried out using our periodic NEO-DFT implementation within the all-electron FHI-aims code \cite{xu2022nuclear, blum2009ab, abbott2025roadmap}.
For the electronic subsystem, we employed the PBE \cite{perdew1996generalized} exchange-correlation (XC) functional with the intermediate numeric atom-centered orbitals (NAO) basis set and the intermediate numerical integration grid setting \cite{blum2009ab}.
Self-consistent field (SCF) calculations were performed using a 0.0172 eV Fermi smearing (corresponding to the Fermi-Dirac distribution at approximately 200 K) and an 11$\times$11$\times$11 Monkhorst-Pack $k$-point grid for Brillouin zone sampling.
{Due to the presence of van Hove singularities, the electronic DOS was evaluated on a denser 110$\times$110$\times$110 $k$-point grid using the tetrahedron method.
(see Supplemental Materials \cite{supp} for additional convergence tests)}

\par
For the protonic subsystem, we used the Hartree-Fock exact exchange for the XC approximation and the PB4-F2 Gaussian-type basis set \cite{yu2020development}. Using only the exact proton-proton exchange is justified because the proton correlation terms are typically negligible even in molecular systems like H$_2$ where the equilibrium bond length is $\sim$0.74 $\text{\AA}$ as discussed in literature \cite{auer2010localized,pavo2020multicomponent}.
In comparison, in H$_3$S at this high pressure, the nearest H-H distance is $\sim$1.5 \AA.
The electron-proton correlation was approximated using the local density approximation (LDA)-like epc17-2 functional \cite{yang2017development,brorsen2017multicomponent}.
The quantum protons were taken to be in their stationary ground state. 
The Debye temperature of H$_3$S ($\sim$1500~K) is reported to be much higher than $T_c$ \cite{talantsev2020advanced}, implying that hydrogen atoms tend to occupy their quantum-mechanical ground state even at $\sim$200K.
Using the constrained-NEO (CNEO) method \cite{xu2020constrained,xu2020full,liu2025JCP}, we also found that the Feynman-Hibbs centroid potential \cite{feynman_hibbs} shows a single minimum with a rather large force constant for the quantum protons at the midpoint of the S-H-S coordinate. This is also consistent with having all the corresponding phonons being in their lowest eigenstate at 200 K 
(see Supplemental Materials \cite{supp} for details).

{
\par
Additionally, we examine how the quantum nature of the proton modifies the phonons in H$_3$S. 
We calculated the phonon dispersions of H$_3$S at 200~GPa with the finite-displacement method using both standard DFT and NEO-DFT calculations.  
Following previous studies \cite{errea2015high,mishra2025electron}, the calculations were performed with a $4\times4\times4$ supercell and a $6\times6\times6$ Monkhorst-Pack $k$-point grid.  
Because of the high symmetry of the \textit{Im$\bar{3}$m} structure, only two distinct atomic displacements (one hydrogen and one sulfur atom) were required to construct the full Hessian matrix using finite displacements, which was subsequently analyzed with the \textsc{phonopy} package \cite{phonopy-phono3py-JPCM}.
For the standard DFT calculations, we employed the same computational setting as used for the electronic-structure calculations.
The NEO-DFT phonon calculations used identical parameters, except that the displaced hydrogen atom was treated as a quantum proton within the constrained NEO-DFT framework \cite{xu2020constrained,xu2020full,liu2025JCP}, while all other nuclei were treated as classical particles.
}

\section{Results and Discussion}
\label{sec:result}
\subsection{Quantum Protons}

\par 
We start by examining the basic properties of the quantum protons obtained in our NEO-DFT calculations.
The expectation values of the position operator, $ \langle \mathbf{\hat r} \rangle$, for all protons remain centered between the two sulfur atoms; 
the quantum proton treatment
maintains the high-symmetry \textit{Im$\bar{3}$m} phase observed with classical protons. 
In systems with soft potential energy surfaces or reduced lattice symmetry, 
NQEs often lead to spontaneous symmetry breaking or off-center localization of light nuclei \cite{wikfeldt2014communication,cahlik2021significance}.
However, in high-pressure H$_3$S, 
the Feynman-Hibbs centroid potential shows a steep and symmetric single-well potential for quantum protons between two sulfur atoms (see Supplemental Materials \cite{supp}). 
The steep slope of the quantum potential, with the strongly confined proton wavefunctions, effectively suppresses the symmetry-breaking quantum fluctuations, helping to preserve the high symmetry of the lattice.

\begin{table}[b]
    \centering
    \caption{List of calculated proton properties. 
    Spread refers to the spatial spread of the proton probability density in Å$^2$. 
    Protonic eigenvalues and zero-point energies (ZPE) are given in eV. For the calculations with the electron-proton correlation (epc), the epc17-2 approximation was used.  }
    \label{tab:proton}
    \begin{tabularx}{\columnwidth}{XXXX}
        \hline\hline
              & Spread  & Eigenvalue   & ZPE  \\
        \hline
        H$_3$S w/o epc        & 0.00895        & $-21.734$     & 1.149    \\
        H$_3$S w/ epc        & 0.02002        & $-21.645$     & 0.389    \\
        D$_3$S w/o epc        & 0.00617        & $-22.339$     & 0.843    \\
        \hline\hline
    \end{tabularx}
\end{table}

\begin{figure*}[t]
    \centering
    \includegraphics[width=0.96\textwidth]{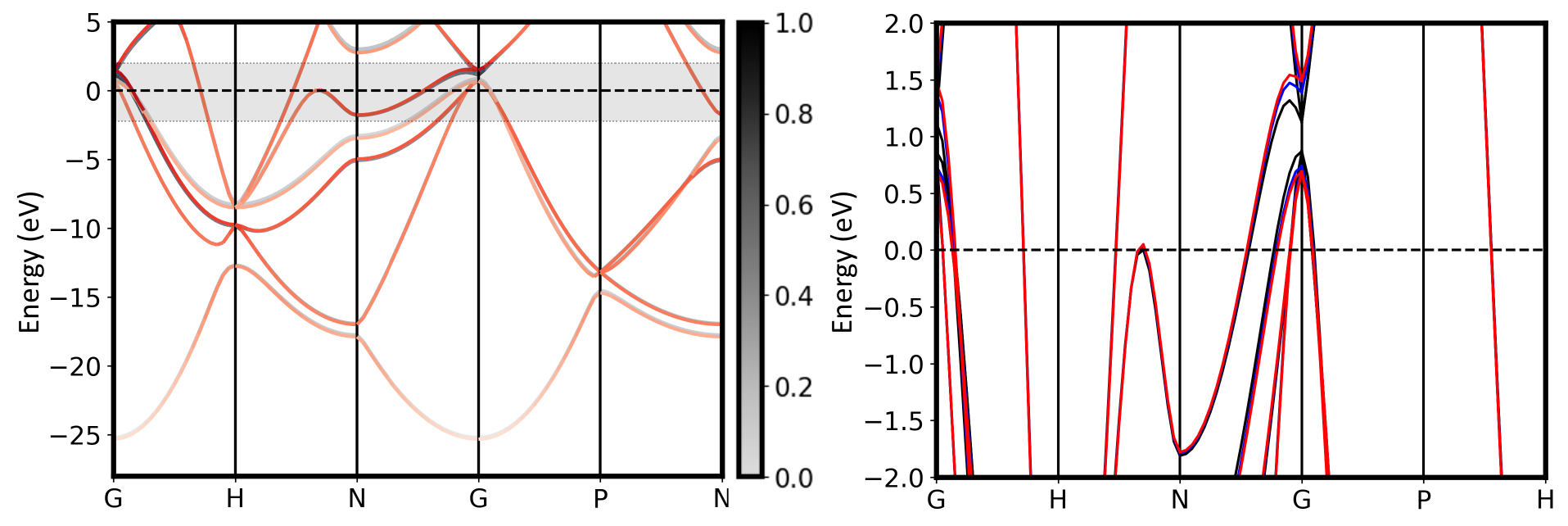}
    \caption{Left panel: Species-projected electronic band structures of H$_3$S at 200 GPa, calculated with the 11$\times$11$\times$11 Monkhorst-Pack $k$-point grid using standard DFT (black), NEO-DFT with deuterium (blue), and hydrogen (red). 
    The color scale indicates the contributions from hydrogen or deuterium atoms to the electronic states.
    A van Hove singularity is visible along the high symmetry path from $H$ to $N$.
    Right panel: Enlarged view of the band structure in the shaded region near the Fermi energy, from -2 eV to 2 eV.
    }
    \label{fig:BS}
\end{figure*}

Table \ref{tab:proton} summarizes spatial spreads and eigenvalues of quantum H/D nuclei as well as the zero-point energies.  Let us first discuss the results without the electron-proton correlation included. 
The spread of the quantum H/D orbital is defined as $\langle r^2\rangle- \langle\mathbf{\hat r}\rangle^2$, and their small values indicate strong localization of quantum nuclei within the crystal lattice.
This observation further justifies the $\Gamma$-point-only calculation for quantum nuclei.
As expected, the lighter hydrogen nuclei exhibit larger spreads of 0.009 \AA$^2$ compared to 0.006 \AA$^2$ for deuterium, reflecting their greater quantum delocalization.
These results reflect the stronger quantum effects of hydrogen relative to deuterium, consistent with its smaller mass and greater nuclear delocalization.
Because they are quite localized and sufficiently far from each other, the three protonic Kohn-Sham states are degenerate with an eigenvalue of -21.734 eV in H$_3$S.
For deuterium, the eigenvalues are lowered to -22.339 eV. 

Taking into account the quantum-mechanical correlation among electrons and protons further modifies the properties of quantum nuclei. 
As noted above, we used the particular LDA-type approximation, epc17-2 \cite{yang2017development}, for the electron-proton correlation (epc) functional in the multicomponent DFT.
As shown in Table \ref{tab:proton}, the electron-proton correlation increases the spatial spread of the proton density from 0.009 \AA$^2$ to 0.020 \AA$^2$ and slightly raises the protonic Kohn-Sham eigenvalues from $-$21.734 eV to $-$21.645 eV.
This behavior reflects that the electron-proton correlation slightly softens the effective potential experienced by the protons and leads to their delocalization.
Table \ref{tab:proton} also lists the zero-point energy (ZPE), defined here as the total energy difference between standard DFT and NEO-DFT calculations, divided by the number of quantum nuclei, $\Delta E/n^p$.
As previously discussed for NEO calculations \cite{pavo2020multicomponent,tao2021analytical}, the epc functional is essential for an accurate estimate of ZPE, yielding 0.389 eV per proton. 
Despite its importance for ZPE, we did not observe meaningful differences in the band structure and DOS between the calculations with and without the epc functional contribution (see Supplemental Materials \cite{supp} for details). 
This shows that, for H$_3$S at 200 GPa, the electron-proton correlation has only minor effects on electronic structure.
For the deuterium, the epc functional would need to be derived differently from that for proton, 
and the larger mass of deuterium leads to significantly reduced quantum delocalization, making electron-nuclear quantum correlation effects even less important.
Therefore, the electron-proton correlation is not included in the calculations for discussing the electronic structure in the following section.

\subsection{Electronic structure}

\begin{figure}[t]
    \centering
    \includegraphics[width=0.48\textwidth]{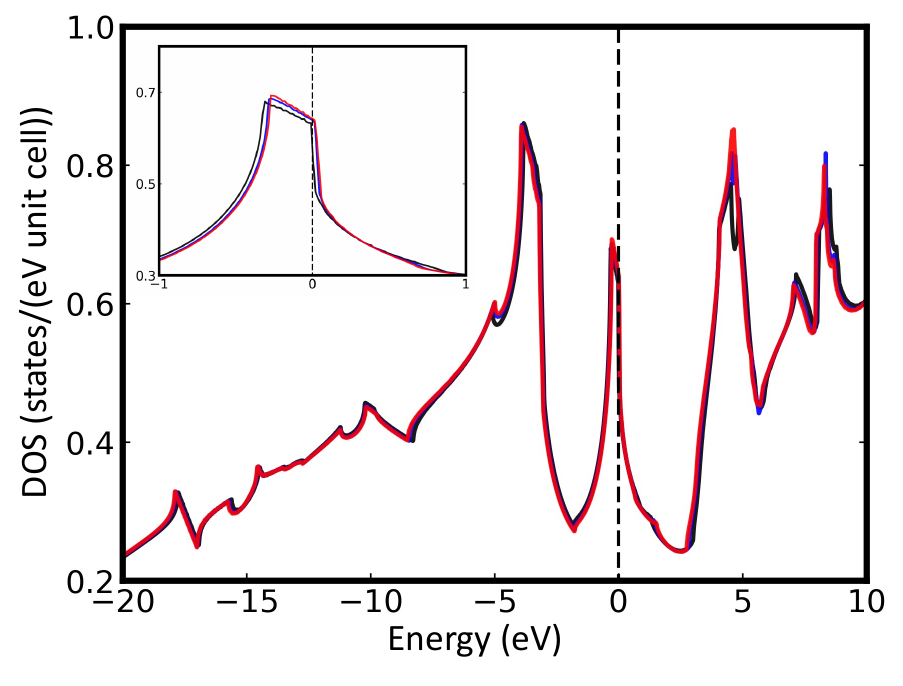}
    \caption{{Electronic DOS of H$_3$S at 200 GPa, calculated using a 110$\times$110$\times$110 Monkhorst-Pack $k$-point grid using standard DFT (black), NEO-DFT with deuterium (blue), and hydrogen (red).
    The inset shows an enlarged view of van Hove singularities near the Fermi energy.}}
    \label{fig:DOS}
\end{figure}

\begin{figure*}[t]
    \centering
    \includegraphics[width=0.96\textwidth]{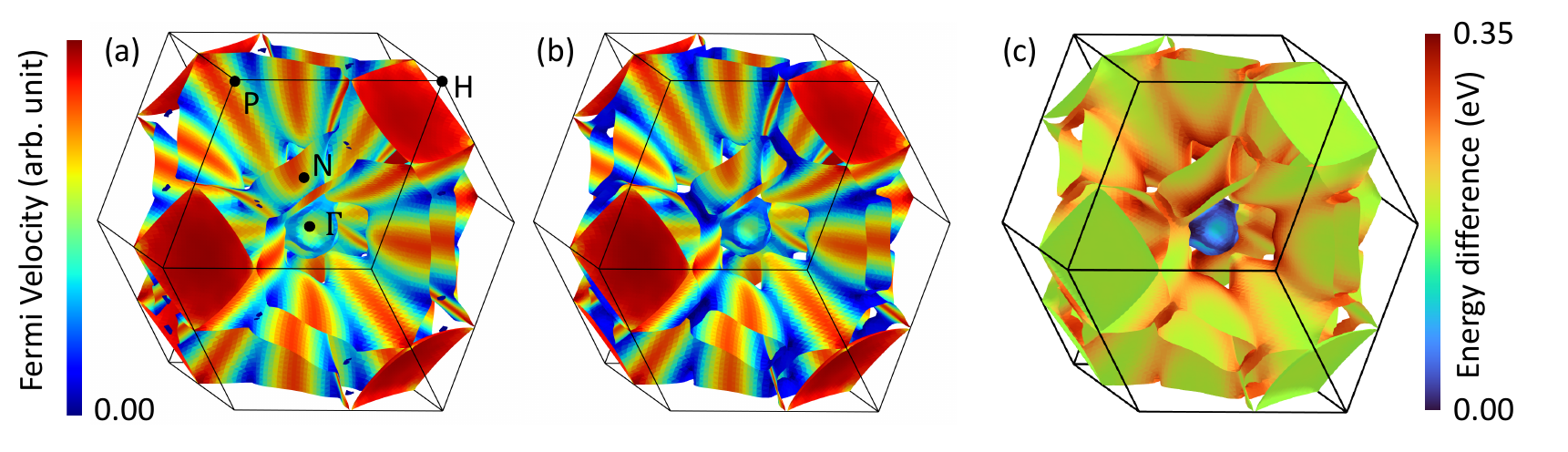}
    \caption{Fermi surface of H$_3$S at 200 GPa. Panels compare (a) standard DFT and (b) NEO-DFT with quantum protons. Blue regions indicate Fermi velocities approaching zero ($d\epsilon/d\bf{k} \approx 0$). The panel (c) shows the energy difference between NEO-DFT and standard DFT, $\epsilon_\text{NEO-DFT}({\bf k})-\epsilon_\text{DFT}({\bf k})$, on the NEO-DFT Fermi surface. Red regions in panel (c) indicate larger energy shifts caused by proton NQEs. }
    \label{fig:FS}
\end{figure*}

\par 
The calculated band structures of H$_3$S are shown in Fig. \ref{fig:BS}.
{All the band structures obtained from standard DFT as well as from NEO-DFT calculations for H$_3$S and D$_3$S are individually shown for detailed analysis in the Supplemental Materials \cite{supp}.}
The results agree with existing works of the \textit{Im$\bar{3}$m} phase at 200 GPa \cite{duan2014pressure,bernstein2015superconducts,quan2016van}.
The species-projected bands show significant hydrogen character in the energy window from -10 eV to 5 eV, as indicated by the color scale.
In this energy range, the electronic states exhibit strong hybridization between sulfur and hydrogen, with hydrogen contributions reaching up to 70\%.
A van Hove singularity is observed near the Fermi energy, in agreement with its proposed role in enhancing the electron-phonon coupling and thus high $T_c$ of H$_3$S as discussed in Ref. \citenum{quan2016van}.

\par 
Nuclear quantum effects of protons introduce subtle modifications to the electronic band structure in the NEO-DFT calculations.
The right panel of Fig. \ref{fig:BS} shows an enlarged view of the band structure near the Fermi energy from -2 to 2 eV.
The NQE-induced changes are particularly evident around the high-symmetry $\Gamma$ point, where hydrogen contributions dominate.
Notably, the vertex at the $\Gamma$ point at around 1 eV above the Fermi energy becomes more pronounced (i.e., from black lines to red lines) when NQEs are included, although the BCS critical temperature would not be affected appreciably by such changes far from the Fermi energy.
The bands near the van Hove singularity, seen between $H$ and $N$ high-symmetry points, are particularly relevant in the context of BCS superconductivity. 
{They are shifted upward in energy when NQEs are taken into account, but only very slightly, which is consistent with the finding in the earlier work using a perturbative approach \cite{sano2016effect}.}
For deuterium (blue lines), these changes are less pronounced compared to hydrogen, reflecting its reduced quantum delocalization due to the larger nuclear mass.

\par 
The corresponding electronic density of states (DOS) is shown in Fig.~\ref{fig:DOS}.
The overall shape of DOS agrees well with existing DFT studies from the literature \cite{bernstein2015superconducts,papaconstantopoulos2015cubic,quan2016van}.
Fig. \ref{fig:DOS} also shows a zoomed-in view of the DOS near the Fermi energy, $\epsilon_F$. 
In the vicinity of $\epsilon_F$, one can observe two distinct features associated with the two van Hove singularities \cite{quan2016van} at approximately 0.0 eV and $-$0.3 eV.
With NQEs, the DOS slightly moves toward higher energy, and
the DOS near $\epsilon_F$ increases in the NEO-DFT calculation.
For the deuterium, the changes in DOS are qualitatively similar but smaller in magnitude, as expected.
{This observation is consistent with the previous work employing the perturbative treatment of NQEs on the electronic structure  \cite{sano2016effect}.
At the same time, while Ref. \citenum{sano2016effect} reported reshaping of the DOS near the Fermi level, the changes observed here primarily manifest as small shifts.} 
As a result of NQEs, the integrated DOS within $\pm$0.1 eV of the Fermi level, $N(\epsilon_F)$, increases by approximately 6.9\% for hydrogen and 5.0\% for deuterium.
This enhancement corresponds to an estimated {2.5\% to 3.4\%} increase in the superconducting critical temperature $T_c$ under the Allen-Dynes limit of the strong coupling.
For the typical $T_c$ of $\sim$200 K reported for H$_3$S \cite{du2025superconducting}, these NQE-induced DOS changes would translate to an increase in $T_c$ of only approximately 5 K. 
This difference is significantly smaller than
uncertainties that stem from other approximations typically adopted in calculating the critical temperature from first-principles theory, like the constant-DOS approximation and the semi-empirical Coulomb pseudopotential.

\par 
Figure \ref{fig:FS} (a) and (b) show the calculated Fermi surface of H$_3$S from the standard DFT calculation and the NEO-DFT calculation with quantum protons, respectively. 
The Fermi surfaces exhibit multiple interconnected sheets and pockets that are characteristic of the metallic phase.
Blue regions correspond to areas where the Fermi velocity, $d\epsilon/d\mathbf{k}$, approaches zero, highlighting the saddle points associated with van Hove singularities.
While the overall topology of the Fermi surface remains similar, NQEs slightly blue-shift the band energies, causing one saddle point along the $N$-$H$ path to shift further from the Fermi energy and the other one to move closer.
Figure \ref{fig:FS} (c) shows the energy shifts induced by NQEs, $\epsilon_\text{NEO-DFT}({\bf k})-\epsilon_\text{DFT}({\bf k})$, on the NEO-DFT Fermi surface ${\bf k}_\text{NEO-DFT}$.
The pocket near $\Gamma$ remains largely unchanged, but regions of small Fermi velocity (blue color areas in Fig. \ref{fig:FS} (b)) are especially affected, as indicated by the appearance of the red regions in Fig. \ref{fig:FS} (c).
This observation underscores the influence of NQEs on the van Hove singularities and their associated peaks in the DOS, although we found the DOS changes are rather minimal in terms of the corresponding $T_c$ changes.

\subsection{Phonons}

\begin{figure}[th]
    \centering
    \includegraphics[width=0.48\textwidth]{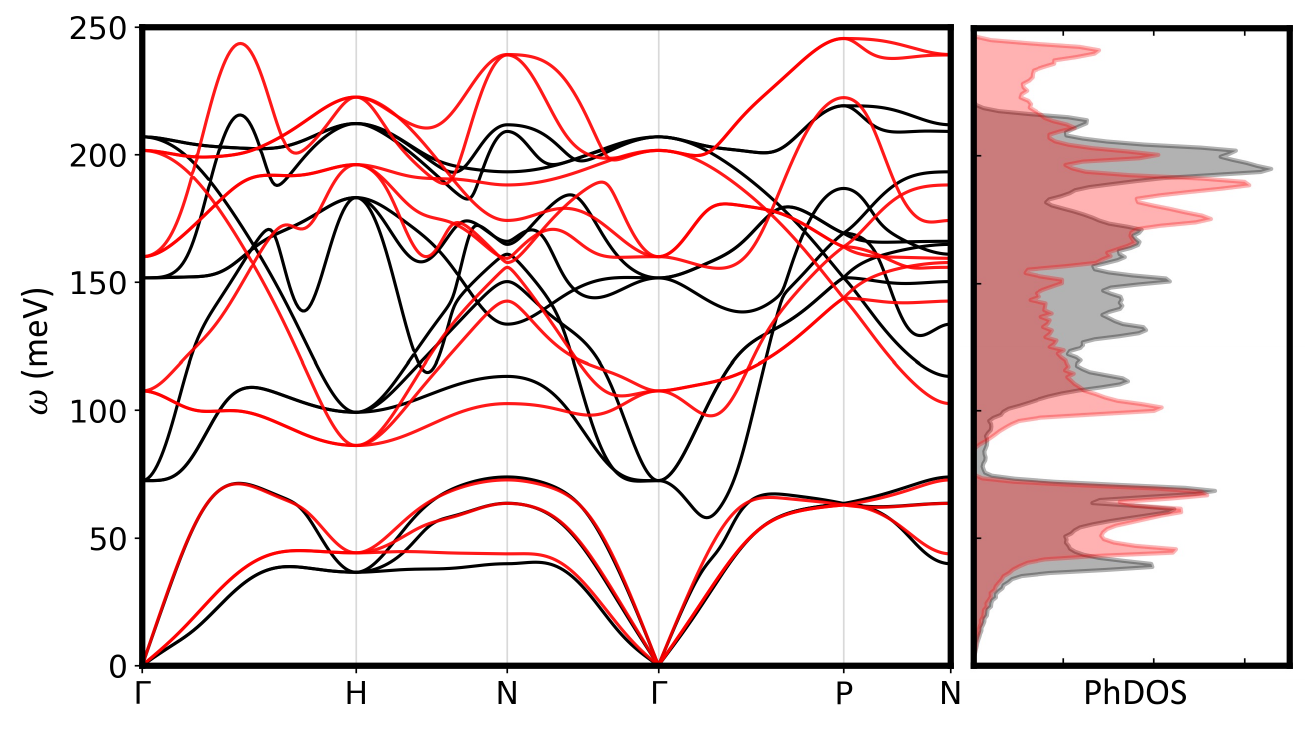}
    \caption{
    Left Panel: Phonon dispersion of H$_3$S at 200~GPa obtained from standard DFT (black) and NEO-DFT (red) calculations using a $4\times4\times4$ supercell and a $6\times6\times6$ $k$-point grid.
    Right Panel: Corresponding phonon density of states, calculated with 1~meV Gaussian broadening.}
    \label{fig:phonon}
\end{figure}

{
The phonon dispersions and phonon density of states of H$_3$S are shown in Fig.~\ref{fig:phonon}.  
The comparison between the standard DFT (black) and NEO-DFT results (red) closely resembles the harmonic phonon spectrum and the phonon spectrum with the quantum anharmonic correction \cite{errea2013first,zacharias2020theory,zacharias2023anharmonic} in the literature for H$_3$S at comparable pressures~ \cite{errea2015high,mishra2025electron}.
{Note that our NEO-DFT calculations include the proton quantum effects through a self-consistently defined effective quantum potential \cite{liu2025JCP}} without relying on the Born-Oppenheimer approximation, 
which is inherently assumed when the nuclear quantum effects are treated separately from the underlying electronic structure calculations. 
Despite these differences in approximations involved in our NEO-DFT, the stochastic self-consistent harmonic approximation (SSCHA) \cite{errea2013first}, and the anharmonic special displacement method (A-SDM) \cite{zacharias2020theory}, all calculations yield remarkably similar phonon spectra.
{Unlike the modest impact of NQEs on the electronic structure near the Fermi level, phonon calculations including NQEs all lead to a distinct modification of the high-frequency hydrogen-dominated phonons, while the low-frequency acoustic branches associated with sulfur atoms remain largely unchanged.} 

{This behavior reflects the dominant role of zero-point motion of the hydrogen-derived modes in the \textit{Im$\bar{3}$m} phase}; quantum fluctuations of the light hydrogen atoms effectively increase the curvature of the potential energy surface along the S-H bonds, shifting the corresponding phonon frequencies to higher energies, while the S-H bond-bending modes are softened.  
\citeauthor{errea2015high} demonstrated that this phonon renormalization reduces the electron-phonon coupling constant $\lambda$ by approximately 30\%, which in turn lowers $T_c$ by 20-25\%, 
whereas the overall spectral shape of the phonon density of states retains its characteristic shape \cite{errea2015high}.  
A similar conclusion was also reached in a recent work by \citeauthor{mishra2025electron}, who further incorporated electron-phonon vertex and non-adiabatic corrections and found that 
these effects, together with nuclear quantum motion,
are essential to reproduce the experimental $T_c$ of H$_3$S \cite{mishra2025electron}. 
{Our NEO-DFT calculations also support these same trends, confirming that an explicit quantum treatment of the protons captures the key nuclear quantum contributions of the lattice dynamics in high-pressure H$_3$S.}
}

\section{Conclusion}
\label{sec:conclusion}
\par 
Experimental measurements and confirmation of superconductivity under extreme conditions remain challenging, 
and first-principles theory plays a unique role in providing both predictive insights and interpretive frameworks.
The discovery of superconducting high-pressure H$_3$S represents a great success of the modern first-principles calculation approach for predicting the phonon-mediated BCS superconductivity.
As the field continues to advance, further assessment and refinement of existing approximations are desirable for increasingly more accurate predictions.

In this work, we investigated the nuclear quantum effects on electronic structure-derived properties of H$_3$S in the \textit{Im$\bar{3}$m} high-pressure phase, a particularly relevant question given the presence of unique S-H-S bridging bonds.  
Specifically, using the NEO-DFT method within a multicomponent Kohn-Sham framework, protons are modeled quantum mechanically on an equal footing as electrons in the first-principles calculations.
We examined how the quantum mechanical nature of protons influences the electronic density of states (DOS), the Fermi surface, and phonons for their role in phonon-mediated BCS superconductivity.
Our work showed that proton quantum effects introduce only modest modifications to the electronic band structure near the Fermi energy, 
resulting in a slight increase in DOS at the Fermi energy and small shifts of van Hove singularities. 
These changes are estimated to impact the prediction of the critical temperature by only a couple of percent, approximately 5 K for H$_3$S in the Allen-Dynes' strong coupling limit. 

On the other hand, the quantum-renormalized phonon dispersion, obtained from the multicomponent DFT, shows that the quantum/anharmonicity-induced hardening of hydrogen-dominated phonons is essential for an accurate prediction of the critical temperature, arriving at the same conclusion as previous studies that used separate methods for computing the quantum anharmonic corrections \cite{errea2015high,mishra2025electron}.
While nuclear quantum effects exert only a minor influence on the electronic band structure, their dominant contribution to the isotope dependence of $T_c$ arises through the phonon-related properties, particularly through the high-frequency H-S stretching phonon and the associated electron-phonon coupling strength.
The evaluation of the electron-phonon coupling within the NEO-DFT method is complicated and is beyond the scope of the present work and will be the subject of future investigation.

In summary, this study establishes that the NEO-DFT method provides a unified first-principles theoretical framework for treating coupled electron-proton quantum effects, enabling simultaneous access to both electronic and phonon properties to predict phonon-mediated BCS superconductivity in hydrogen-rich materials.


\begin{acknowledgments}
The work is partly funded by the National Science Foundation, under No. OAC-2209858.
The Research Computing at University of North Carolina at Chapel Hill is acknowledged for providing computational resources. 
\end{acknowledgments}


\begin{thebibliography}{59}%
\makeatletter
\providecommand \@ifxundefined [1]{%
 \@ifx{#1\undefined}
}%
\providecommand \@ifnum [1]{%
 \ifnum #1\expandafter \@firstoftwo
 \else \expandafter \@secondoftwo
 \fi
}%
\providecommand \@ifx [1]{%
 \ifx #1\expandafter \@firstoftwo
 \else \expandafter \@secondoftwo
 \fi
}%
\providecommand \natexlab [1]{#1}%
\providecommand \enquote  [1]{``#1''}%
\providecommand \bibnamefont  [1]{#1}%
\providecommand \bibfnamefont [1]{#1}%
\providecommand \citenamefont [1]{#1}%
\providecommand \href@noop [0]{\@secondoftwo}%
\providecommand \href [0]{\begingroup \@sanitize@url \@href}%
\providecommand \@href[1]{\@@startlink{#1}\@@href}%
\providecommand \@@href[1]{\endgroup#1\@@endlink}%
\providecommand \@sanitize@url [0]{\catcode `\\12\catcode `\$12\catcode `\&12\catcode `\#12\catcode `\^12\catcode `\_12\catcode `\%12\relax}%
\providecommand \@@startlink[1]{}%
\providecommand \@@endlink[0]{}%
\providecommand \url  [0]{\begingroup\@sanitize@url \@url }%
\providecommand \@url [1]{\endgroup\@href {#1}{\urlprefix }}%
\providecommand \urlprefix  [0]{URL }%
\providecommand \Eprint [0]{\href }%
\providecommand \doibase [0]{https://doi.org/}%
\providecommand \selectlanguage [0]{\@gobble}%
\providecommand \bibinfo  [0]{\@secondoftwo}%
\providecommand \bibfield  [0]{\@secondoftwo}%
\providecommand \translation [1]{[#1]}%
\providecommand \BibitemOpen [0]{}%
\providecommand \bibitemStop [0]{}%
\providecommand \bibitemNoStop [0]{.\EOS\space}%
\providecommand \EOS [0]{\spacefactor3000\relax}%
\providecommand \BibitemShut  [1]{\csname bibitem#1\endcsname}%
\let\auto@bib@innerbib\@empty
\bibitem [{\citenamefont {Calandra}\ \emph {et~al.}(2007)\citenamefont {Calandra}, \citenamefont {Lazzeri},\ and\ \citenamefont {Mauri}}]{calandra2007anharmonic}%
  \BibitemOpen
  \bibfield  {author} {\bibinfo {author} {\bibfnamefont {M.}~\bibnamefont {Calandra}}, \bibinfo {author} {\bibfnamefont {M.}~\bibnamefont {Lazzeri}},\ and\ \bibinfo {author} {\bibfnamefont {F.}~\bibnamefont {Mauri}},\ }\bibfield  {title} {\bibinfo {title} {Anharmonic and non-adiabatic effects in mgb2: Implications for the isotope effect and interpretation of raman spectra},\ }\href@noop {} {\bibfield  {journal} {\bibinfo  {journal} {Phys. C, Supercond.}\ }\textbf {\bibinfo {volume} {456}},\ \bibinfo {pages} {38} (\bibinfo {year} {2007})}\BibitemShut {NoStop}%
\bibitem [{\citenamefont {Errea}(2016)}]{errea2016approaching}%
  \BibitemOpen
  \bibfield  {author} {\bibinfo {author} {\bibfnamefont {I.}~\bibnamefont {Errea}},\ }\bibfield  {title} {\bibinfo {title} {Approaching the strongly anharmonic limit with ab initio calculations of materials’ vibrational properties--a colloquium},\ }\href@noop {} {\bibfield  {journal} {\bibinfo  {journal} {Eur. Phys. J. B.}\ }\textbf {\bibinfo {volume} {89}},\ \bibinfo {pages} {237} (\bibinfo {year} {2016})}\BibitemShut {NoStop}%
\bibitem [{\citenamefont {Giustino}(2017)}]{giustino2017electron}%
  \BibitemOpen
  \bibfield  {author} {\bibinfo {author} {\bibfnamefont {F.}~\bibnamefont {Giustino}},\ }\bibfield  {title} {\bibinfo {title} {Electron-phonon interactions from first principles},\ }\href@noop {} {\bibfield  {journal} {\bibinfo  {journal} {Rev. Mod. Phys.}\ }\textbf {\bibinfo {volume} {89}},\ \bibinfo {pages} {015003} (\bibinfo {year} {2017})}\BibitemShut {NoStop}%
\bibitem [{\citenamefont {Zurek}\ and\ \citenamefont {Bi}(2019)}]{zurek2019high}%
  \BibitemOpen
  \bibfield  {author} {\bibinfo {author} {\bibfnamefont {E.}~\bibnamefont {Zurek}}\ and\ \bibinfo {author} {\bibfnamefont {T.}~\bibnamefont {Bi}},\ }\bibfield  {title} {\bibinfo {title} {High-temperature superconductivity in alkaline and rare earth polyhydrides at high pressure: A theoretical perspective},\ }\href@noop {} {\bibfield  {journal} {\bibinfo  {journal} {J. Chem. Phys.}\ }\textbf {\bibinfo {volume} {150}},\ \bibinfo {pages} {050901} (\bibinfo {year} {2019})}\BibitemShut {NoStop}%
\bibitem [{\citenamefont {Flores-Livas}\ \emph {et~al.}(2020)\citenamefont {Flores-Livas}, \citenamefont {Boeri}, \citenamefont {Sanna}, \citenamefont {Profeta}, \citenamefont {Arita},\ and\ \citenamefont {Eremets}}]{flores2020perspective}%
  \BibitemOpen
  \bibfield  {author} {\bibinfo {author} {\bibfnamefont {J.~A.}\ \bibnamefont {Flores-Livas}}, \bibinfo {author} {\bibfnamefont {L.}~\bibnamefont {Boeri}}, \bibinfo {author} {\bibfnamefont {A.}~\bibnamefont {Sanna}}, \bibinfo {author} {\bibfnamefont {G.}~\bibnamefont {Profeta}}, \bibinfo {author} {\bibfnamefont {R.}~\bibnamefont {Arita}},\ and\ \bibinfo {author} {\bibfnamefont {M.}~\bibnamefont {Eremets}},\ }\bibfield  {title} {\bibinfo {title} {A perspective on conventional high-temperature superconductors at high pressure: Methods and materials},\ }\href@noop {} {\bibfield  {journal} {\bibinfo  {journal} {Phys. Rep.}\ }\textbf {\bibinfo {volume} {856}},\ \bibinfo {pages} {1} (\bibinfo {year} {2020})}\BibitemShut {NoStop}%
\bibitem [{\citenamefont {Bardeen}\ \emph {et~al.}(1957)\citenamefont {Bardeen}, \citenamefont {Cooper},\ and\ \citenamefont {Schrieffer}}]{bardeen1957theory}%
  \BibitemOpen
  \bibfield  {author} {\bibinfo {author} {\bibfnamefont {J.}~\bibnamefont {Bardeen}}, \bibinfo {author} {\bibfnamefont {L.~N.}\ \bibnamefont {Cooper}},\ and\ \bibinfo {author} {\bibfnamefont {J.~R.}\ \bibnamefont {Schrieffer}},\ }\bibfield  {title} {\bibinfo {title} {Theory of superconductivity},\ }\href@noop {} {\bibfield  {journal} {\bibinfo  {journal} {Phys. Rev.}\ }\textbf {\bibinfo {volume} {108}},\ \bibinfo {pages} {1175} (\bibinfo {year} {1957})}\BibitemShut {NoStop}%
\bibitem [{\citenamefont {Lee}\ \emph {et~al.}(2023)\citenamefont {Lee}, \citenamefont {Ponc{\'e}}, \citenamefont {Bushick}, \citenamefont {Hajinazar}, \citenamefont {Lafuente-Bartolome}, \citenamefont {Leveillee}, \citenamefont {Lian}, \citenamefont {Lihm}, \citenamefont {Macheda}, \citenamefont {Mori} \emph {et~al.}}]{lee2023electron}%
  \BibitemOpen
  \bibfield  {author} {\bibinfo {author} {\bibfnamefont {H.}~\bibnamefont {Lee}}, \bibinfo {author} {\bibfnamefont {S.}~\bibnamefont {Ponc{\'e}}}, \bibinfo {author} {\bibfnamefont {K.}~\bibnamefont {Bushick}}, \bibinfo {author} {\bibfnamefont {S.}~\bibnamefont {Hajinazar}}, \bibinfo {author} {\bibfnamefont {J.}~\bibnamefont {Lafuente-Bartolome}}, \bibinfo {author} {\bibfnamefont {J.}~\bibnamefont {Leveillee}}, \bibinfo {author} {\bibfnamefont {C.}~\bibnamefont {Lian}}, \bibinfo {author} {\bibfnamefont {J.-M.}\ \bibnamefont {Lihm}}, \bibinfo {author} {\bibfnamefont {F.}~\bibnamefont {Macheda}}, \bibinfo {author} {\bibfnamefont {H.}~\bibnamefont {Mori}}, \emph {et~al.},\ }\bibfield  {title} {\bibinfo {title} {Electron--phonon physics from first principles using the epw code},\ }\href@noop {} {\bibfield  {journal} {\bibinfo  {journal} {Npj Comput. Mater.}\ }\textbf {\bibinfo {volume} {9}},\ \bibinfo {pages} {156} (\bibinfo {year} {2023})}\BibitemShut {NoStop}%
\bibitem [{\citenamefont {Mishra}\ \emph {et~al.}(2025)\citenamefont {Mishra}, \citenamefont {Mori},\ and\ \citenamefont {Margine}}]{mishra2025electron}%
  \BibitemOpen
  \bibfield  {author} {\bibinfo {author} {\bibfnamefont {S.~B.}\ \bibnamefont {Mishra}}, \bibinfo {author} {\bibfnamefont {H.}~\bibnamefont {Mori}},\ and\ \bibinfo {author} {\bibfnamefont {E.~R.}\ \bibnamefont {Margine}},\ }\bibfield  {title} {\bibinfo {title} {Electron-phonon vertex correction effect in superconducting h3s},\ }\href@noop {} {\bibfield  {journal} {\bibinfo  {journal} {arXiv preprint arXiv:2507.01897}\ } (\bibinfo {year} {2025})}\BibitemShut {NoStop}%
\bibitem [{\citenamefont {Pellegrini}\ and\ \citenamefont {Sanna}(2024)}]{pellegrini2024ab}%
  \BibitemOpen
  \bibfield  {author} {\bibinfo {author} {\bibfnamefont {C.}~\bibnamefont {Pellegrini}}\ and\ \bibinfo {author} {\bibfnamefont {A.}~\bibnamefont {Sanna}},\ }\bibfield  {title} {\bibinfo {title} {Ab initio methods for superconductivity},\ }\href@noop {} {\bibfield  {journal} {\bibinfo  {journal} {Nat. Rev. Phys.}\ }\textbf {\bibinfo {volume} {6}},\ \bibinfo {pages} {509} (\bibinfo {year} {2024})}\BibitemShut {NoStop}%
\bibitem [{\citenamefont {Ashcroft}(2004)}]{ashcroft2004hydrogen}%
  \BibitemOpen
  \bibfield  {author} {\bibinfo {author} {\bibfnamefont {N.}~\bibnamefont {Ashcroft}},\ }\bibfield  {title} {\bibinfo {title} {Hydrogen dominant metallic alloys: high temperature superconductors?},\ }\href@noop {} {\bibfield  {journal} {\bibinfo  {journal} {Phys. Rev. Lett.}\ }\textbf {\bibinfo {volume} {92}},\ \bibinfo {pages} {187002} (\bibinfo {year} {2004})}\BibitemShut {NoStop}%
\bibitem [{\citenamefont {Duan}\ \emph {et~al.}(2014)\citenamefont {Duan}, \citenamefont {Liu}, \citenamefont {Tian}, \citenamefont {Li}, \citenamefont {Huang}, \citenamefont {Zhao}, \citenamefont {Yu}, \citenamefont {Liu}, \citenamefont {Tian},\ and\ \citenamefont {Cui}}]{duan2014pressure}%
  \BibitemOpen
  \bibfield  {author} {\bibinfo {author} {\bibfnamefont {D.}~\bibnamefont {Duan}}, \bibinfo {author} {\bibfnamefont {Y.}~\bibnamefont {Liu}}, \bibinfo {author} {\bibfnamefont {F.}~\bibnamefont {Tian}}, \bibinfo {author} {\bibfnamefont {D.}~\bibnamefont {Li}}, \bibinfo {author} {\bibfnamefont {X.}~\bibnamefont {Huang}}, \bibinfo {author} {\bibfnamefont {Z.}~\bibnamefont {Zhao}}, \bibinfo {author} {\bibfnamefont {H.}~\bibnamefont {Yu}}, \bibinfo {author} {\bibfnamefont {B.}~\bibnamefont {Liu}}, \bibinfo {author} {\bibfnamefont {W.}~\bibnamefont {Tian}},\ and\ \bibinfo {author} {\bibfnamefont {T.}~\bibnamefont {Cui}},\ }\bibfield  {title} {\bibinfo {title} {Pressure-induced metallization of dense (h2s) 2h2 with high-t c superconductivity},\ }\href@noop {} {\bibfield  {journal} {\bibinfo  {journal} {Sci. Rep.}\ }\textbf {\bibinfo {volume} {4}},\ \bibinfo {pages} {6968} (\bibinfo {year} {2014})}\BibitemShut {NoStop}%
\bibitem [{\citenamefont {Drozdov}\ \emph {et~al.}(2015)\citenamefont {Drozdov}, \citenamefont {Eremets}, \citenamefont {Troyan}, \citenamefont {Ksenofontov},\ and\ \citenamefont {Shylin}}]{drozdov2015conventional}%
  \BibitemOpen
  \bibfield  {author} {\bibinfo {author} {\bibfnamefont {A.}~\bibnamefont {Drozdov}}, \bibinfo {author} {\bibfnamefont {M.}~\bibnamefont {Eremets}}, \bibinfo {author} {\bibfnamefont {I.}~\bibnamefont {Troyan}}, \bibinfo {author} {\bibfnamefont {V.}~\bibnamefont {Ksenofontov}},\ and\ \bibinfo {author} {\bibfnamefont {S.~I.}\ \bibnamefont {Shylin}},\ }\bibfield  {title} {\bibinfo {title} {Conventional superconductivity at 203 kelvin at high pressures in the sulfur hydride system},\ }\href@noop {} {\bibfield  {journal} {\bibinfo  {journal} {Nature}\ }\textbf {\bibinfo {volume} {525}},\ \bibinfo {pages} {73} (\bibinfo {year} {2015})}\BibitemShut {NoStop}%
\bibitem [{\citenamefont {Du}\ \emph {et~al.}(2025)\citenamefont {Du}, \citenamefont {Drozdov}, \citenamefont {Minkov}, \citenamefont {Balakirev}, \citenamefont {Kong}, \citenamefont {Smith}, \citenamefont {Yan}, \citenamefont {Shen}, \citenamefont {Gegenwart},\ and\ \citenamefont {Eremets}}]{du2025superconducting}%
  \BibitemOpen
  \bibfield  {author} {\bibinfo {author} {\bibfnamefont {F.}~\bibnamefont {Du}}, \bibinfo {author} {\bibfnamefont {A.~P.}\ \bibnamefont {Drozdov}}, \bibinfo {author} {\bibfnamefont {V.~S.}\ \bibnamefont {Minkov}}, \bibinfo {author} {\bibfnamefont {F.~F.}\ \bibnamefont {Balakirev}}, \bibinfo {author} {\bibfnamefont {P.}~\bibnamefont {Kong}}, \bibinfo {author} {\bibfnamefont {G.~A.}\ \bibnamefont {Smith}}, \bibinfo {author} {\bibfnamefont {J.}~\bibnamefont {Yan}}, \bibinfo {author} {\bibfnamefont {B.}~\bibnamefont {Shen}}, \bibinfo {author} {\bibfnamefont {P.}~\bibnamefont {Gegenwart}},\ and\ \bibinfo {author} {\bibfnamefont {M.~I.}\ \bibnamefont {Eremets}},\ }\bibfield  {title} {\bibinfo {title} {Superconducting gap of h3s measured by tunnelling spectroscopy},\ }\href@noop {} {\bibfield  {journal} {\bibinfo  {journal} {Nature}\ }\textbf {\bibinfo {volume} {641}},\ \bibinfo {pages} {619} (\bibinfo {year} {2025})}\BibitemShut {NoStop}%
\bibitem [{\citenamefont {Errea}\ \emph {et~al.}(2015)\citenamefont {Errea}, \citenamefont {Calandra}, \citenamefont {Pickard}, \citenamefont {Nelson}, \citenamefont {Needs}, \citenamefont {Li}, \citenamefont {Liu}, \citenamefont {Zhang}, \citenamefont {Ma},\ and\ \citenamefont {Mauri}}]{errea2015high}%
  \BibitemOpen
  \bibfield  {author} {\bibinfo {author} {\bibfnamefont {I.}~\bibnamefont {Errea}}, \bibinfo {author} {\bibfnamefont {M.}~\bibnamefont {Calandra}}, \bibinfo {author} {\bibfnamefont {C.~J.}\ \bibnamefont {Pickard}}, \bibinfo {author} {\bibfnamefont {J.}~\bibnamefont {Nelson}}, \bibinfo {author} {\bibfnamefont {R.~J.}\ \bibnamefont {Needs}}, \bibinfo {author} {\bibfnamefont {Y.}~\bibnamefont {Li}}, \bibinfo {author} {\bibfnamefont {H.}~\bibnamefont {Liu}}, \bibinfo {author} {\bibfnamefont {Y.}~\bibnamefont {Zhang}}, \bibinfo {author} {\bibfnamefont {Y.}~\bibnamefont {Ma}},\ and\ \bibinfo {author} {\bibfnamefont {F.}~\bibnamefont {Mauri}},\ }\bibfield  {title} {\bibinfo {title} {High-pressure hydrogen sulfide from first principles: a strongly anharmonic phonon-mediated superconductor},\ }\href@noop {} {\bibfield  {journal} {\bibinfo  {journal} {Phys. Rev. Lett.}\ }\textbf {\bibinfo {volume} {114}},\ \bibinfo {pages} {157004} (\bibinfo {year} {2015})}\BibitemShut {NoStop}%
\bibitem [{\citenamefont {Carbotte}(1990)}]{carbotte1990properties}%
  \BibitemOpen
  \bibfield  {author} {\bibinfo {author} {\bibfnamefont {J.}~\bibnamefont {Carbotte}},\ }\bibfield  {title} {\bibinfo {title} {Properties of boson-exchange superconductors},\ }\href@noop {} {\bibfield  {journal} {\bibinfo  {journal} {Rev. Mod. Phys.}\ }\textbf {\bibinfo {volume} {62}},\ \bibinfo {pages} {1027} (\bibinfo {year} {1990})}\BibitemShut {NoStop}%
\bibitem [{\citenamefont {Bernstein}\ \emph {et~al.}(2015)\citenamefont {Bernstein}, \citenamefont {Hellberg}, \citenamefont {Johannes}, \citenamefont {Mazin},\ and\ \citenamefont {Mehl}}]{bernstein2015superconducts}%
  \BibitemOpen
  \bibfield  {author} {\bibinfo {author} {\bibfnamefont {N.}~\bibnamefont {Bernstein}}, \bibinfo {author} {\bibfnamefont {C.~S.}\ \bibnamefont {Hellberg}}, \bibinfo {author} {\bibfnamefont {M.}~\bibnamefont {Johannes}}, \bibinfo {author} {\bibfnamefont {I.}~\bibnamefont {Mazin}},\ and\ \bibinfo {author} {\bibfnamefont {M.}~\bibnamefont {Mehl}},\ }\bibfield  {title} {\bibinfo {title} {What superconducts in sulfur hydrides under pressure and why},\ }\href@noop {} {\bibfield  {journal} {\bibinfo  {journal} {Phys. Rev. B}\ }\textbf {\bibinfo {volume} {91}},\ \bibinfo {pages} {060511} (\bibinfo {year} {2015})}\BibitemShut {NoStop}%
\bibitem [{\citenamefont {Papaconstantopoulos}\ \emph {et~al.}(2015)\citenamefont {Papaconstantopoulos}, \citenamefont {Klein}, \citenamefont {Mehl},\ and\ \citenamefont {Pickett}}]{papaconstantopoulos2015cubic}%
  \BibitemOpen
  \bibfield  {author} {\bibinfo {author} {\bibfnamefont {D.}~\bibnamefont {Papaconstantopoulos}}, \bibinfo {author} {\bibfnamefont {B.~M.}\ \bibnamefont {Klein}}, \bibinfo {author} {\bibfnamefont {M.}~\bibnamefont {Mehl}},\ and\ \bibinfo {author} {\bibfnamefont {W.}~\bibnamefont {Pickett}},\ }\bibfield  {title} {\bibinfo {title} {Cubic h 3 s around 200 gpa: An atomic hydrogen superconductor stabilized by sulfur},\ }\href@noop {} {\bibfield  {journal} {\bibinfo  {journal} {Phys. Rev. B}\ }\textbf {\bibinfo {volume} {91}},\ \bibinfo {pages} {184511} (\bibinfo {year} {2015})}\BibitemShut {NoStop}%
\bibitem [{\citenamefont {Quan}\ and\ \citenamefont {Pickett}(2016)}]{quan2016van}%
  \BibitemOpen
  \bibfield  {author} {\bibinfo {author} {\bibfnamefont {Y.}~\bibnamefont {Quan}}\ and\ \bibinfo {author} {\bibfnamefont {W.~E.}\ \bibnamefont {Pickett}},\ }\bibfield  {title} {\bibinfo {title} {Van hove singularities and spectral smearing in high-temperature superconducting h 3 s},\ }\href@noop {} {\bibfield  {journal} {\bibinfo  {journal} {Phys. Rev. B}\ }\textbf {\bibinfo {volume} {93}},\ \bibinfo {pages} {104526} (\bibinfo {year} {2016})}\BibitemShut {NoStop}%
\bibitem [{\citenamefont {Ortenzi}\ \emph {et~al.}(2016)\citenamefont {Ortenzi}, \citenamefont {Cappelluti},\ and\ \citenamefont {Pietronero}}]{ortenzi2016band}%
  \BibitemOpen
  \bibfield  {author} {\bibinfo {author} {\bibfnamefont {L.}~\bibnamefont {Ortenzi}}, \bibinfo {author} {\bibfnamefont {E.}~\bibnamefont {Cappelluti}},\ and\ \bibinfo {author} {\bibfnamefont {L.}~\bibnamefont {Pietronero}},\ }\bibfield  {title} {\bibinfo {title} {Band structure and electron-phonon coupling in h 3 s: A tight-binding model},\ }\href@noop {} {\bibfield  {journal} {\bibinfo  {journal} {Phys. Rev. B}\ }\textbf {\bibinfo {volume} {94}},\ \bibinfo {pages} {064507} (\bibinfo {year} {2016})}\BibitemShut {NoStop}%
\bibitem [{\citenamefont {Durajski}\ and\ \citenamefont {Szczk{e}{\'s}niak}(2017)}]{durajski2017first}%
  \BibitemOpen
  \bibfield  {author} {\bibinfo {author} {\bibfnamefont {A.~P.}\ \bibnamefont {Durajski}}\ and\ \bibinfo {author} {\bibfnamefont {R.}~\bibnamefont {Szczk{e}{\'s}niak}},\ }\bibfield  {title} {\bibinfo {title} {First-principles study of superconducting hydrogen sulfide at pressure up to 500 gpa},\ }\href@noop {} {\bibfield  {journal} {\bibinfo  {journal} {Sci. Rep.}\ }\textbf {\bibinfo {volume} {7}},\ \bibinfo {pages} {4473} (\bibinfo {year} {2017})}\BibitemShut {NoStop}%
\bibitem [{\citenamefont {Szczk{e}{\'s}niak}\ and\ \citenamefont {Durajski}(2018)}]{szczkesniak2018unusual}%
  \BibitemOpen
  \bibfield  {author} {\bibinfo {author} {\bibfnamefont {R.}~\bibnamefont {Szczk{e}{\'s}niak}}\ and\ \bibinfo {author} {\bibfnamefont {A.~P.}\ \bibnamefont {Durajski}},\ }\bibfield  {title} {\bibinfo {title} {Unusual sulfur isotope effect and extremely high critical temperature in h3s superconductor},\ }\href@noop {} {\bibfield  {journal} {\bibinfo  {journal} {Sci. Rep.}\ }\textbf {\bibinfo {volume} {8}},\ \bibinfo {pages} {6037} (\bibinfo {year} {2018})}\BibitemShut {NoStop}%
\bibitem [{\citenamefont {Jarlborg}\ and\ \citenamefont {Bianconi}(2016)}]{jarlborg2016breakdown}%
  \BibitemOpen
  \bibfield  {author} {\bibinfo {author} {\bibfnamefont {T.}~\bibnamefont {Jarlborg}}\ and\ \bibinfo {author} {\bibfnamefont {A.}~\bibnamefont {Bianconi}},\ }\bibfield  {title} {\bibinfo {title} {Breakdown of the migdal approximation at lifshitz transitions with giant zero-point motion in the h3s superconductor},\ }\href@noop {} {\bibfield  {journal} {\bibinfo  {journal} {Sci. Rep.}\ }\textbf {\bibinfo {volume} {6}},\ \bibinfo {pages} {24816} (\bibinfo {year} {2016})}\BibitemShut {NoStop}%
\bibitem [{\citenamefont {Errea}\ \emph {et~al.}(2016)\citenamefont {Errea}, \citenamefont {Calandra}, \citenamefont {Pickard}, \citenamefont {Nelson}, \citenamefont {Needs}, \citenamefont {Li}, \citenamefont {Liu}, \citenamefont {Zhang}, \citenamefont {Ma},\ and\ \citenamefont {Mauri}}]{errea2016quantum}%
  \BibitemOpen
  \bibfield  {author} {\bibinfo {author} {\bibfnamefont {I.}~\bibnamefont {Errea}}, \bibinfo {author} {\bibfnamefont {M.}~\bibnamefont {Calandra}}, \bibinfo {author} {\bibfnamefont {C.~J.}\ \bibnamefont {Pickard}}, \bibinfo {author} {\bibfnamefont {J.~R.}\ \bibnamefont {Nelson}}, \bibinfo {author} {\bibfnamefont {R.~J.}\ \bibnamefont {Needs}}, \bibinfo {author} {\bibfnamefont {Y.}~\bibnamefont {Li}}, \bibinfo {author} {\bibfnamefont {H.}~\bibnamefont {Liu}}, \bibinfo {author} {\bibfnamefont {Y.}~\bibnamefont {Zhang}}, \bibinfo {author} {\bibfnamefont {Y.}~\bibnamefont {Ma}},\ and\ \bibinfo {author} {\bibfnamefont {F.}~\bibnamefont {Mauri}},\ }\bibfield  {title} {\bibinfo {title} {Quantum hydrogen-bond symmetrization in the superconducting hydrogen sulfide system},\ }\href@noop {} {\bibfield  {journal} {\bibinfo  {journal} {Nature}\ }\textbf {\bibinfo {volume} {532}},\ \bibinfo {pages} {81} (\bibinfo {year} {2016})}\BibitemShut {NoStop}%
\bibitem [{\citenamefont {Sano}\ \emph {et~al.}(2016)\citenamefont {Sano}, \citenamefont {Koretsune}, \citenamefont {Tadano}, \citenamefont {Akashi},\ and\ \citenamefont {Arita}}]{sano2016effect}%
  \BibitemOpen
  \bibfield  {author} {\bibinfo {author} {\bibfnamefont {W.}~\bibnamefont {Sano}}, \bibinfo {author} {\bibfnamefont {T.}~\bibnamefont {Koretsune}}, \bibinfo {author} {\bibfnamefont {T.}~\bibnamefont {Tadano}}, \bibinfo {author} {\bibfnamefont {R.}~\bibnamefont {Akashi}},\ and\ \bibinfo {author} {\bibfnamefont {R.}~\bibnamefont {Arita}},\ }\bibfield  {title} {\bibinfo {title} {Effect of van hove singularities on high-t c superconductivity in h 3 s},\ }\href@noop {} {\bibfield  {journal} {\bibinfo  {journal} {Phys. Rev. B}\ }\textbf {\bibinfo {volume} {93}},\ \bibinfo {pages} {094525} (\bibinfo {year} {2016})}\BibitemShut {NoStop}%
\bibitem [{\citenamefont {Taureau}\ \emph {et~al.}(2024)\citenamefont {Taureau}, \citenamefont {Cherubini}, \citenamefont {Morresi},\ and\ \citenamefont {Casula}}]{taureau2024quantum}%
  \BibitemOpen
  \bibfield  {author} {\bibinfo {author} {\bibfnamefont {R.}~\bibnamefont {Taureau}}, \bibinfo {author} {\bibfnamefont {M.}~\bibnamefont {Cherubini}}, \bibinfo {author} {\bibfnamefont {T.}~\bibnamefont {Morresi}},\ and\ \bibinfo {author} {\bibfnamefont {M.}~\bibnamefont {Casula}},\ }\bibfield  {title} {\bibinfo {title} {Quantum symmetrization transition in superconducting sulfur hydride from quantum monte carlo and path integral molecular dynamics},\ }\href@noop {} {\bibfield  {journal} {\bibinfo  {journal} {Npj Comput. Mater.}\ }\textbf {\bibinfo {volume} {10}},\ \bibinfo {pages} {56} (\bibinfo {year} {2024})}\BibitemShut {NoStop}%
\bibitem [{\citenamefont {Xu}\ \emph {et~al.}(2022)\citenamefont {Xu}, \citenamefont {Zhou}, \citenamefont {Tao}, \citenamefont {Malbon}, \citenamefont {Blum}, \citenamefont {Hammes-Schiffer},\ and\ \citenamefont {Kanai}}]{xu2022nuclear}%
  \BibitemOpen
  \bibfield  {author} {\bibinfo {author} {\bibfnamefont {J.}~\bibnamefont {Xu}}, \bibinfo {author} {\bibfnamefont {R.}~\bibnamefont {Zhou}}, \bibinfo {author} {\bibfnamefont {Z.}~\bibnamefont {Tao}}, \bibinfo {author} {\bibfnamefont {C.}~\bibnamefont {Malbon}}, \bibinfo {author} {\bibfnamefont {V.}~\bibnamefont {Blum}}, \bibinfo {author} {\bibfnamefont {S.}~\bibnamefont {Hammes-Schiffer}},\ and\ \bibinfo {author} {\bibfnamefont {Y.}~\bibnamefont {Kanai}},\ }\bibfield  {title} {\bibinfo {title} {Nuclear--electronic orbital approach to quantization of protons in periodic electronic structure calculations},\ }\href@noop {} {\bibfield  {journal} {\bibinfo  {journal} {J. Chem. Phys.}\ }\textbf {\bibinfo {volume} {156}},\ \bibinfo {pages} {224111} (\bibinfo {year} {2022})}\BibitemShut {NoStop}%
\bibitem [{\citenamefont {Cai}\ \emph {et~al.}(2024)\citenamefont {Cai}, \citenamefont {Li}, \citenamefont {Fu}, \citenamefont {Yang}, \citenamefont {Mei}, \citenamefont {Nie}, \citenamefont {Li}, \citenamefont {Liu}, \citenamefont {Ke}, \citenamefont {Wang} \emph {et~al.}}]{cai2024deuteration}%
  \BibitemOpen
  \bibfield  {author} {\bibinfo {author} {\bibfnamefont {G.}~\bibnamefont {Cai}}, \bibinfo {author} {\bibfnamefont {Y.}~\bibnamefont {Li}}, \bibinfo {author} {\bibfnamefont {Y.}~\bibnamefont {Fu}}, \bibinfo {author} {\bibfnamefont {H.}~\bibnamefont {Yang}}, \bibinfo {author} {\bibfnamefont {L.}~\bibnamefont {Mei}}, \bibinfo {author} {\bibfnamefont {Z.}~\bibnamefont {Nie}}, \bibinfo {author} {\bibfnamefont {T.}~\bibnamefont {Li}}, \bibinfo {author} {\bibfnamefont {H.}~\bibnamefont {Liu}}, \bibinfo {author} {\bibfnamefont {Y.}~\bibnamefont {Ke}}, \bibinfo {author} {\bibfnamefont {X.-L.}\ \bibnamefont {Wang}}, \emph {et~al.},\ }\bibfield  {title} {\bibinfo {title} {Deuteration-enhanced neutron contrasts to probe amorphous domain sizes in organic photovoltaic bulk heterojunction films},\ }\href@noop {} {\bibfield  {journal} {\bibinfo  {journal} {Nat. Commun.}\ }\textbf {\bibinfo {volume} {15}},\ \bibinfo {pages} {2784} (\bibinfo {year} {2024})}\BibitemShut {NoStop}%
\bibitem [{\citenamefont {Akashi}\ \emph {et~al.}(2015)\citenamefont {Akashi}, \citenamefont {Kawamura}, \citenamefont {Tsuneyuki}, \citenamefont {Nomura},\ and\ \citenamefont {Arita}}]{akashi2015first}%
  \BibitemOpen
  \bibfield  {author} {\bibinfo {author} {\bibfnamefont {R.}~\bibnamefont {Akashi}}, \bibinfo {author} {\bibfnamefont {M.}~\bibnamefont {Kawamura}}, \bibinfo {author} {\bibfnamefont {S.}~\bibnamefont {Tsuneyuki}}, \bibinfo {author} {\bibfnamefont {Y.}~\bibnamefont {Nomura}},\ and\ \bibinfo {author} {\bibfnamefont {R.}~\bibnamefont {Arita}},\ }\bibfield  {title} {\bibinfo {title} {First-principles study of the pressure and crystal-structure dependences of the superconducting transition temperature in compressed sulfur hydrides},\ }\href@noop {} {\bibfield  {journal} {\bibinfo  {journal} {Phys. Rev. B}\ }\textbf {\bibinfo {volume} {91}},\ \bibinfo {pages} {224513} (\bibinfo {year} {2015})}\BibitemShut {NoStop}%
\bibitem [{\citenamefont {Allen}\ and\ \citenamefont {Dynes}(1975)}]{allen1975transition}%
  \BibitemOpen
  \bibfield  {author} {\bibinfo {author} {\bibfnamefont {P.~B.}\ \bibnamefont {Allen}}\ and\ \bibinfo {author} {\bibfnamefont {R.}~\bibnamefont {Dynes}},\ }\bibfield  {title} {\bibinfo {title} {Transition temperature of strong-coupled superconductors reanalyzed},\ }\href@noop {} {\bibfield  {journal} {\bibinfo  {journal} {Phys. Rev. B}\ }\textbf {\bibinfo {volume} {12}},\ \bibinfo {pages} {905} (\bibinfo {year} {1975})}\BibitemShut {NoStop}%
\bibitem [{\citenamefont {Capitani}\ \emph {et~al.}(1982)\citenamefont {Capitani}, \citenamefont {Nalewajski},\ and\ \citenamefont {Parr}}]{capitani1982non}%
  \BibitemOpen
  \bibfield  {author} {\bibinfo {author} {\bibfnamefont {J.~F.}\ \bibnamefont {Capitani}}, \bibinfo {author} {\bibfnamefont {R.~F.}\ \bibnamefont {Nalewajski}},\ and\ \bibinfo {author} {\bibfnamefont {R.~G.}\ \bibnamefont {Parr}},\ }\bibfield  {title} {\bibinfo {title} {Non-born--oppenheimer density functional theory of molecular systems},\ }\href@noop {} {\bibfield  {journal} {\bibinfo  {journal} {J. Chem. Phys.}\ }\textbf {\bibinfo {volume} {76}},\ \bibinfo {pages} {568} (\bibinfo {year} {1982})}\BibitemShut {NoStop}%
\bibitem [{\citenamefont {Kreibich}\ and\ \citenamefont {Gross}(2001)}]{kreibich2001multicomponent}%
  \BibitemOpen
  \bibfield  {author} {\bibinfo {author} {\bibfnamefont {T.}~\bibnamefont {Kreibich}}\ and\ \bibinfo {author} {\bibfnamefont {E.}~\bibnamefont {Gross}},\ }\bibfield  {title} {\bibinfo {title} {Multicomponent density-functional theory for electrons and nuclei},\ }\href@noop {} {\bibfield  {journal} {\bibinfo  {journal} {Phys. Rev. Lett.}\ }\textbf {\bibinfo {volume} {86}},\ \bibinfo {pages} {2984} (\bibinfo {year} {2001})}\BibitemShut {NoStop}%
\bibitem [{\citenamefont {Webb}\ \emph {et~al.}(2002)\citenamefont {Webb}, \citenamefont {Iordanov},\ and\ \citenamefont {Hammes-Schiffer}}]{webb_multiconfigurational_2002}%
  \BibitemOpen
  \bibfield  {author} {\bibinfo {author} {\bibfnamefont {S.~P.}\ \bibnamefont {Webb}}, \bibinfo {author} {\bibfnamefont {T.}~\bibnamefont {Iordanov}},\ and\ \bibinfo {author} {\bibfnamefont {S.}~\bibnamefont {Hammes-Schiffer}},\ }\bibfield  {title} {\bibinfo {title} {Multiconfigurational nuclear-electronic orbital approach: {Incorporation} of nuclear quantum effects in electronic structure calculations},\ }\href {https://doi.org/10.1063/1.1494980} {\bibfield  {journal} {\bibinfo  {journal} {J. Chem. Phys.}\ }\textbf {\bibinfo {volume} {117}},\ \bibinfo {pages} {4106} (\bibinfo {year} {2002})}\BibitemShut {NoStop}%
\bibitem [{\citenamefont {Hammes-Schiffer}(2021)}]{hammes-schiffer_nuclearelectronic_2021}%
  \BibitemOpen
  \bibfield  {author} {\bibinfo {author} {\bibfnamefont {S.}~\bibnamefont {Hammes-Schiffer}},\ }\bibfield  {title} {\bibinfo {title} {Nuclear–electronic orbital methods: {Foundations} and prospects},\ }\href {https://doi.org/10.1063/5.0053576} {\bibfield  {journal} {\bibinfo  {journal} {J. Chem. Phys.}\ }\textbf {\bibinfo {volume} {155}},\ \bibinfo {pages} {030901} (\bibinfo {year} {2021})}\BibitemShut {NoStop}%
\bibitem [{\citenamefont {Pak}\ \emph {et~al.}(2007)\citenamefont {Pak}, \citenamefont {Chakraborty},\ and\ \citenamefont {Hammes-Schiffer}}]{pak_density_2007}%
  \BibitemOpen
  \bibfield  {author} {\bibinfo {author} {\bibfnamefont {M.~V.}\ \bibnamefont {Pak}}, \bibinfo {author} {\bibfnamefont {A.}~\bibnamefont {Chakraborty}},\ and\ \bibinfo {author} {\bibfnamefont {S.}~\bibnamefont {Hammes-Schiffer}},\ }\bibfield  {title} {\bibinfo {title} {Density {Functional} {Theory} {Treatment} of {Electron} {Correlation} in the {Nuclear}-{Electronic} {Orbital} {Approach}},\ }\href {https://doi.org/10.1021/jp0704463} {\bibfield  {journal} {\bibinfo  {journal} {J. Phys. Chem. A}\ }\textbf {\bibinfo {volume} {111}},\ \bibinfo {pages} {4522} (\bibinfo {year} {2007})}\BibitemShut {NoStop}%
\bibitem [{\citenamefont {Chakraborty}\ \emph {et~al.}(2008)\citenamefont {Chakraborty}, \citenamefont {Pak},\ and\ \citenamefont {Hammes-Schiffer}}]{chakraborty_development_2008}%
  \BibitemOpen
  \bibfield  {author} {\bibinfo {author} {\bibfnamefont {A.}~\bibnamefont {Chakraborty}}, \bibinfo {author} {\bibfnamefont {M.~V.}\ \bibnamefont {Pak}},\ and\ \bibinfo {author} {\bibfnamefont {S.}~\bibnamefont {Hammes-Schiffer}},\ }\bibfield  {title} {\bibinfo {title} {Development of {Electron}-{Proton} {Density} {Functionals} for {Multicomponent} {Density} {Functional} {Theory}},\ }\href {https://doi.org/10.1103/PhysRevLett.101.153001} {\bibfield  {journal} {\bibinfo  {journal} {Phys. Rev. Lett.}\ }\textbf {\bibinfo {volume} {101}},\ \bibinfo {pages} {153001} (\bibinfo {year} {2008})}\BibitemShut {NoStop}%
\bibitem [{\citenamefont {Udagawa}\ and\ \citenamefont {Tachikawa}(2006)}]{udagawa2006h}%
  \BibitemOpen
  \bibfield  {author} {\bibinfo {author} {\bibfnamefont {T.}~\bibnamefont {Udagawa}}\ and\ \bibinfo {author} {\bibfnamefont {M.}~\bibnamefont {Tachikawa}},\ }\bibfield  {title} {\bibinfo {title} {H/ d isotope effect on porphine and porphycene molecules with multicomponent hybrid density functional theory},\ }\href@noop {} {\bibfield  {journal} {\bibinfo  {journal} {J. Chem. Phys.}\ }\textbf {\bibinfo {volume} {125}} (\bibinfo {year} {2006})}\BibitemShut {NoStop}%
\bibitem [{\citenamefont {Pavosevic}\ \emph {et~al.}(2020)\citenamefont {Pavosevic}, \citenamefont {Culpitt},\ and\ \citenamefont {Hammes-Schiffer}}]{pavo2020multicomponent}%
  \BibitemOpen
  \bibfield  {author} {\bibinfo {author} {\bibfnamefont {F.}~\bibnamefont {Pavosevic}}, \bibinfo {author} {\bibfnamefont {T.}~\bibnamefont {Culpitt}},\ and\ \bibinfo {author} {\bibfnamefont {S.}~\bibnamefont {Hammes-Schiffer}},\ }\bibfield  {title} {\bibinfo {title} {Multicomponent quantum chemistry: Integrating electronic and nuclear quantum effects via the nuclear--electronic orbital method},\ }\href@noop {} {\bibfield  {journal} {\bibinfo  {journal} {Chem. Rev.}\ }\textbf {\bibinfo {volume} {120}},\ \bibinfo {pages} {4222} (\bibinfo {year} {2020})}\BibitemShut {NoStop}%
\bibitem [{\citenamefont {Brorsen}\ \emph {et~al.}(2017)\citenamefont {Brorsen}, \citenamefont {Yang},\ and\ \citenamefont {Hammes-Schiffer}}]{brorsen2017multicomponent}%
  \BibitemOpen
  \bibfield  {author} {\bibinfo {author} {\bibfnamefont {K.~R.}\ \bibnamefont {Brorsen}}, \bibinfo {author} {\bibfnamefont {Y.}~\bibnamefont {Yang}},\ and\ \bibinfo {author} {\bibfnamefont {S.}~\bibnamefont {Hammes-Schiffer}},\ }\bibfield  {title} {\bibinfo {title} {Multicomponent density functional theory: Impact of nuclear quantum effects on proton affinities and geometries},\ }\href@noop {} {\bibfield  {journal} {\bibinfo  {journal} {J. Phys. Chem. Lett.}\ }\textbf {\bibinfo {volume} {8}},\ \bibinfo {pages} {3488} (\bibinfo {year} {2017})}\BibitemShut {NoStop}%
\bibitem [{\citenamefont {Yang}\ \emph {et~al.}(2017)\citenamefont {Yang}, \citenamefont {Brorsen}, \citenamefont {Culpitt}, \citenamefont {Pak},\ and\ \citenamefont {Hammes-Schiffer}}]{yang2017development}%
  \BibitemOpen
  \bibfield  {author} {\bibinfo {author} {\bibfnamefont {Y.}~\bibnamefont {Yang}}, \bibinfo {author} {\bibfnamefont {K.~R.}\ \bibnamefont {Brorsen}}, \bibinfo {author} {\bibfnamefont {T.}~\bibnamefont {Culpitt}}, \bibinfo {author} {\bibfnamefont {M.~V.}\ \bibnamefont {Pak}},\ and\ \bibinfo {author} {\bibfnamefont {S.}~\bibnamefont {Hammes-Schiffer}},\ }\bibfield  {title} {\bibinfo {title} {Development of a practical multicomponent density functional for electron-proton correlation to produce accurate proton densities},\ }\href@noop {} {\bibfield  {journal} {\bibinfo  {journal} {J. Chem. Phys.}\ }\textbf {\bibinfo {volume} {147}},\ \bibinfo {pages} {114113} (\bibinfo {year} {2017})}\BibitemShut {NoStop}%
\bibitem [{\citenamefont {Brorsen}\ \emph {et~al.}(2018)\citenamefont {Brorsen}, \citenamefont {Schneider},\ and\ \citenamefont {Hammes-Schiffer}}]{brorsen2018alternative}%
  \BibitemOpen
  \bibfield  {author} {\bibinfo {author} {\bibfnamefont {K.~R.}\ \bibnamefont {Brorsen}}, \bibinfo {author} {\bibfnamefont {P.~E.}\ \bibnamefont {Schneider}},\ and\ \bibinfo {author} {\bibfnamefont {S.}~\bibnamefont {Hammes-Schiffer}},\ }\bibfield  {title} {\bibinfo {title} {Alternative forms and transferability of electron-proton correlation functionals in nuclear-electronic orbital density functional theory},\ }\href@noop {} {\bibfield  {journal} {\bibinfo  {journal} {J. Chem. Phys.}\ }\textbf {\bibinfo {volume} {149}},\ \bibinfo {pages} {044110} (\bibinfo {year} {2018})}\BibitemShut {NoStop}%
\bibitem [{\citenamefont {Tao}\ \emph {et~al.}(2019)\citenamefont {Tao}, \citenamefont {Yang},\ and\ \citenamefont {Hammes-Schiffer}}]{tao2019multicomponent}%
  \BibitemOpen
  \bibfield  {author} {\bibinfo {author} {\bibfnamefont {Z.}~\bibnamefont {Tao}}, \bibinfo {author} {\bibfnamefont {Y.}~\bibnamefont {Yang}},\ and\ \bibinfo {author} {\bibfnamefont {S.}~\bibnamefont {Hammes-Schiffer}},\ }\bibfield  {title} {\bibinfo {title} {Multicomponent density functional theory: Including the density gradient in the electron-proton correlation functional for hydrogen and deuterium},\ }\href@noop {} {\bibfield  {journal} {\bibinfo  {journal} {J. Chem. Phys.}\ }\textbf {\bibinfo {volume} {151}},\ \bibinfo {pages} {124102} (\bibinfo {year} {2019})}\BibitemShut {NoStop}%
\bibitem [{\citenamefont {Blum}\ \emph {et~al.}(2009)\citenamefont {Blum}, \citenamefont {Gehrke}, \citenamefont {Hanke}, \citenamefont {Havu}, \citenamefont {Havu}, \citenamefont {Ren}, \citenamefont {Reuter},\ and\ \citenamefont {Scheffler}}]{blum2009ab}%
  \BibitemOpen
  \bibfield  {author} {\bibinfo {author} {\bibfnamefont {V.}~\bibnamefont {Blum}}, \bibinfo {author} {\bibfnamefont {R.}~\bibnamefont {Gehrke}}, \bibinfo {author} {\bibfnamefont {F.}~\bibnamefont {Hanke}}, \bibinfo {author} {\bibfnamefont {P.}~\bibnamefont {Havu}}, \bibinfo {author} {\bibfnamefont {V.}~\bibnamefont {Havu}}, \bibinfo {author} {\bibfnamefont {X.}~\bibnamefont {Ren}}, \bibinfo {author} {\bibfnamefont {K.}~\bibnamefont {Reuter}},\ and\ \bibinfo {author} {\bibfnamefont {M.}~\bibnamefont {Scheffler}},\ }\bibfield  {title} {\bibinfo {title} {Ab initio molecular simulations with numeric atom-centered orbitals},\ }\href@noop {} {\bibfield  {journal} {\bibinfo  {journal} {Comput. Phys. Commun.}\ }\textbf {\bibinfo {volume} {180}},\ \bibinfo {pages} {2175} (\bibinfo {year} {2009})}\BibitemShut {NoStop}%
\bibitem [{\citenamefont {Abbott}\ \emph {et~al.}(2025)\citenamefont {Abbott}, \citenamefont {Acosta}, \citenamefont {Akkoush}, \citenamefont {Ambrosetti}, \citenamefont {Atalla}, \citenamefont {Bagrets}, \citenamefont {Behler}, \citenamefont {Berger}, \citenamefont {Bieniek}, \citenamefont {Bj{\"o}rk} \emph {et~al.}}]{abbott2025roadmap}%
  \BibitemOpen
  \bibfield  {author} {\bibinfo {author} {\bibfnamefont {J.~W.}\ \bibnamefont {Abbott}}, \bibinfo {author} {\bibfnamefont {C.~M.}\ \bibnamefont {Acosta}}, \bibinfo {author} {\bibfnamefont {A.}~\bibnamefont {Akkoush}}, \bibinfo {author} {\bibfnamefont {A.}~\bibnamefont {Ambrosetti}}, \bibinfo {author} {\bibfnamefont {V.}~\bibnamefont {Atalla}}, \bibinfo {author} {\bibfnamefont {A.}~\bibnamefont {Bagrets}}, \bibinfo {author} {\bibfnamefont {J.}~\bibnamefont {Behler}}, \bibinfo {author} {\bibfnamefont {D.}~\bibnamefont {Berger}}, \bibinfo {author} {\bibfnamefont {B.}~\bibnamefont {Bieniek}}, \bibinfo {author} {\bibfnamefont {J.}~\bibnamefont {Bj{\"o}rk}}, \emph {et~al.},\ }\bibfield  {title} {\bibinfo {title} {Roadmap on advancements of the fhi-aims software package},\ }\href@noop {} {\bibfield  {journal} {\bibinfo  {journal} {arXiv preprint arXiv:2505.00125}\ } (\bibinfo {year} {2025})}\BibitemShut {NoStop}%
\bibitem [{\citenamefont {Perdew}\ \emph {et~al.}(1996)\citenamefont {Perdew}, \citenamefont {Burke},\ and\ \citenamefont {Ernzerhof}}]{perdew1996generalized}%
  \BibitemOpen
  \bibfield  {author} {\bibinfo {author} {\bibfnamefont {J.~P.}\ \bibnamefont {Perdew}}, \bibinfo {author} {\bibfnamefont {K.}~\bibnamefont {Burke}},\ and\ \bibinfo {author} {\bibfnamefont {M.}~\bibnamefont {Ernzerhof}},\ }\bibfield  {title} {\bibinfo {title} {Generalized gradient approximation made simple},\ }\href@noop {} {\bibfield  {journal} {\bibinfo  {journal} {Phys. Rev. Lett.}\ }\textbf {\bibinfo {volume} {77}},\ \bibinfo {pages} {3865} (\bibinfo {year} {1996})}\BibitemShut {NoStop}%
\bibitem [{sup()}]{supp}%
  \BibitemOpen
  \href@noop {} {}\bibinfo {note} {See Supplemental Material at [URL-will-be-inserted-by-publisher] for thermal distribution of protons, individual projected electronic band structures, convergence tests of k-meshes, and effects of electron-proton correlation on electronic structure.}\BibitemShut {Stop}%
\bibitem [{\citenamefont {Yu}\ \emph {et~al.}(2020)\citenamefont {Yu}, \citenamefont {Pavo{\v{s}}evi{\'c}},\ and\ \citenamefont {Hammes-Schiffer}}]{yu2020development}%
  \BibitemOpen
  \bibfield  {author} {\bibinfo {author} {\bibfnamefont {Q.}~\bibnamefont {Yu}}, \bibinfo {author} {\bibfnamefont {F.}~\bibnamefont {Pavo{\v{s}}evi{\'c}}},\ and\ \bibinfo {author} {\bibfnamefont {S.}~\bibnamefont {Hammes-Schiffer}},\ }\bibfield  {title} {\bibinfo {title} {Development of nuclear basis sets for multicomponent quantum chemistry methods},\ }\href@noop {} {\bibfield  {journal} {\bibinfo  {journal} {J. Chem. Phys.}\ }\textbf {\bibinfo {volume} {152}},\ \bibinfo {pages} {244123} (\bibinfo {year} {2020})}\BibitemShut {NoStop}%
\bibitem [{\citenamefont {Auer}\ and\ \citenamefont {Hammes-Schiffer}(2010)}]{auer2010localized}%
  \BibitemOpen
  \bibfield  {author} {\bibinfo {author} {\bibfnamefont {B.}~\bibnamefont {Auer}}\ and\ \bibinfo {author} {\bibfnamefont {S.}~\bibnamefont {Hammes-Schiffer}},\ }\bibfield  {title} {\bibinfo {title} {Localized hartree product treatment of multiple protons in the nuclear-electronic orbital framework},\ }\href@noop {} {\bibfield  {journal} {\bibinfo  {journal} {J. Chem. Phys.}\ }\textbf {\bibinfo {volume} {132}},\ \bibinfo {pages} {084110} (\bibinfo {year} {2010})}\BibitemShut {NoStop}%
\bibitem [{\citenamefont {Talantsev}(2020)}]{talantsev2020advanced}%
  \BibitemOpen
  \bibfield  {author} {\bibinfo {author} {\bibfnamefont {E.}~\bibnamefont {Talantsev}},\ }\bibfield  {title} {\bibinfo {title} {Advanced mcmillan’s equation and its application for the analysis of highly-compressed superconductors},\ }\href@noop {} {\bibfield  {journal} {\bibinfo  {journal} {Supercond. Sci. Technol.}\ }\textbf {\bibinfo {volume} {33}},\ \bibinfo {pages} {094009} (\bibinfo {year} {2020})}\BibitemShut {NoStop}%
\bibitem [{\citenamefont {Xu}\ and\ \citenamefont {Yang}(2020{\natexlab{a}})}]{xu2020constrained}%
  \BibitemOpen
  \bibfield  {author} {\bibinfo {author} {\bibfnamefont {X.}~\bibnamefont {Xu}}\ and\ \bibinfo {author} {\bibfnamefont {Y.}~\bibnamefont {Yang}},\ }\bibfield  {title} {\bibinfo {title} {Constrained nuclear-electronic orbital density functional theory: Energy surfaces with nuclear quantum effects},\ }\href@noop {} {\bibfield  {journal} {\bibinfo  {journal} {J. Chem. Phys.}\ }\textbf {\bibinfo {volume} {152}},\ \bibinfo {pages} {084107} (\bibinfo {year} {2020}{\natexlab{a}})}\BibitemShut {NoStop}%
\bibitem [{\citenamefont {Xu}\ and\ \citenamefont {Yang}(2020{\natexlab{b}})}]{xu2020full}%
  \BibitemOpen
  \bibfield  {author} {\bibinfo {author} {\bibfnamefont {X.}~\bibnamefont {Xu}}\ and\ \bibinfo {author} {\bibfnamefont {Y.}~\bibnamefont {Yang}},\ }\bibfield  {title} {\bibinfo {title} {Full-quantum descriptions of molecular systems from constrained nuclear--electronic orbital density functional theory},\ }\href@noop {} {\bibfield  {journal} {\bibinfo  {journal} {J. Chem. Phys.}\ }\textbf {\bibinfo {volume} {153}},\ \bibinfo {pages} {074106} (\bibinfo {year} {2020}{\natexlab{b}})}\BibitemShut {NoStop}%
\bibitem [{\citenamefont {Liu}\ \emph {et~al.}(2025)\citenamefont {Liu}, \citenamefont {Xu},\ and\ \citenamefont {Kanai}}]{liu2025JCP}%
  \BibitemOpen
  \bibfield  {author} {\bibinfo {author} {\bibfnamefont {S.}~\bibnamefont {Liu}}, \bibinfo {author} {\bibfnamefont {J.}~\bibnamefont {Xu}},\ and\ \bibinfo {author} {\bibfnamefont {Y.}~\bibnamefont {Kanai}},\ }\bibfield  {title} {\bibinfo {title} {Constrained nuclear--electronic orbital method for periodic density functional theory: Application to h2 chemisorption on si (001) surfaces},\ }\href@noop {} {\bibfield  {journal} {\bibinfo  {journal} {J. Chem. Phys.}\ }\textbf {\bibinfo {volume} {163}},\ \bibinfo {pages} {084110} (\bibinfo {year} {2025})}\BibitemShut {NoStop}%
\bibitem [{\citenamefont {Feynman}\ \emph {et~al.}(2010)\citenamefont {Feynman}, \citenamefont {Hibbs},\ and\ \citenamefont {Styer}}]{feynman_hibbs}%
  \BibitemOpen
  \bibfield  {author} {\bibinfo {author} {\bibfnamefont {R.}~\bibnamefont {Feynman}}, \bibinfo {author} {\bibfnamefont {A.}~\bibnamefont {Hibbs}},\ and\ \bibinfo {author} {\bibfnamefont {D.}~\bibnamefont {Styer}},\ }\href {https://books.google.com/books?id=JkMuDAAAQBAJ} {\emph {\bibinfo {title} {Quantum Mechanics and Path Integrals}}},\ Dover Books on Physics\ (\bibinfo  {publisher} {Dover Publications},\ \bibinfo {year} {2010})\BibitemShut {NoStop}%
\bibitem [{\citenamefont {Togo}\ \emph {et~al.}(2023)\citenamefont {Togo}, \citenamefont {Chaput}, \citenamefont {Tadano},\ and\ \citenamefont {Tanaka}}]{phonopy-phono3py-JPCM}%
  \BibitemOpen
  \bibfield  {author} {\bibinfo {author} {\bibfnamefont {A.}~\bibnamefont {Togo}}, \bibinfo {author} {\bibfnamefont {L.}~\bibnamefont {Chaput}}, \bibinfo {author} {\bibfnamefont {T.}~\bibnamefont {Tadano}},\ and\ \bibinfo {author} {\bibfnamefont {I.}~\bibnamefont {Tanaka}},\ }\bibfield  {title} {\bibinfo {title} {Implementation strategies in phonopy and phono3py},\ }\href {https://doi.org/10.1088/1361-648X/acd831} {\bibfield  {journal} {\bibinfo  {journal} {J. Phys. Condens. Matter}\ }\textbf {\bibinfo {volume} {35}},\ \bibinfo {pages} {353001} (\bibinfo {year} {2023})}\BibitemShut {NoStop}%
\bibitem [{\citenamefont {Wikfeldt}\ and\ \citenamefont {Michaelides}(2014)}]{wikfeldt2014communication}%
  \BibitemOpen
  \bibfield  {author} {\bibinfo {author} {\bibfnamefont {K.}~\bibnamefont {Wikfeldt}}\ and\ \bibinfo {author} {\bibfnamefont {A.}~\bibnamefont {Michaelides}},\ }\bibfield  {title} {\bibinfo {title} {Communication: Ab initio simulations of hydrogen-bonded ferroelectrics: Collective tunneling and the origin of geometrical isotope effects},\ }\href@noop {} {\bibfield  {journal} {\bibinfo  {journal} {J. Chem. Phys.}\ }\textbf {\bibinfo {volume} {140}},\ \bibinfo {pages} {041103} (\bibinfo {year} {2014})}\BibitemShut {NoStop}%
\bibitem [{\citenamefont {Cahlik}\ \emph {et~al.}(2021)\citenamefont {Cahlik}, \citenamefont {Hellerstedt}, \citenamefont {Mendieta-Moreno}, \citenamefont {Svec}, \citenamefont {Santhini}, \citenamefont {Pascal}, \citenamefont {Soler-Polo}, \citenamefont {Erlingsson}, \citenamefont {Vyborny}, \citenamefont {Mutombo} \emph {et~al.}}]{cahlik2021significance}%
  \BibitemOpen
  \bibfield  {author} {\bibinfo {author} {\bibfnamefont {A.}~\bibnamefont {Cahlik}}, \bibinfo {author} {\bibfnamefont {J.}~\bibnamefont {Hellerstedt}}, \bibinfo {author} {\bibfnamefont {J.~I.}\ \bibnamefont {Mendieta-Moreno}}, \bibinfo {author} {\bibfnamefont {M.}~\bibnamefont {Svec}}, \bibinfo {author} {\bibfnamefont {V.~M.}\ \bibnamefont {Santhini}}, \bibinfo {author} {\bibfnamefont {S.}~\bibnamefont {Pascal}}, \bibinfo {author} {\bibfnamefont {D.}~\bibnamefont {Soler-Polo}}, \bibinfo {author} {\bibfnamefont {S.~I.}\ \bibnamefont {Erlingsson}}, \bibinfo {author} {\bibfnamefont {K.}~\bibnamefont {Vyborny}}, \bibinfo {author} {\bibfnamefont {P.}~\bibnamefont {Mutombo}}, \emph {et~al.},\ }\bibfield  {title} {\bibinfo {title} {Significance of nuclear quantum effects in hydrogen bonded molecular chains},\ }\href@noop {} {\bibfield  {journal} {\bibinfo  {journal} {ACS nano}\ }\textbf {\bibinfo {volume} {15}},\ \bibinfo {pages} {10357} (\bibinfo {year} {2021})}\BibitemShut {NoStop}%
\bibitem [{\citenamefont {Tao}\ \emph {et~al.}(2021)\citenamefont {Tao}, \citenamefont {Roy}, \citenamefont {Schneider}, \citenamefont {Pavosevic},\ and\ \citenamefont {Hammes-Schiffer}}]{tao2021analytical}%
  \BibitemOpen
  \bibfield  {author} {\bibinfo {author} {\bibfnamefont {Z.}~\bibnamefont {Tao}}, \bibinfo {author} {\bibfnamefont {S.}~\bibnamefont {Roy}}, \bibinfo {author} {\bibfnamefont {P.~E.}\ \bibnamefont {Schneider}}, \bibinfo {author} {\bibfnamefont {F.}~\bibnamefont {Pavosevic}},\ and\ \bibinfo {author} {\bibfnamefont {S.}~\bibnamefont {Hammes-Schiffer}},\ }\bibfield  {title} {\bibinfo {title} {Analytical gradients for nuclear--electronic orbital time-dependent density functional theory: Excited-state geometry optimizations and adiabatic excitation energies},\ }\href@noop {} {\bibfield  {journal} {\bibinfo  {journal} {J. Chem. Theory Comput.}\ }\textbf {\bibinfo {volume} {17}},\ \bibinfo {pages} {5110} (\bibinfo {year} {2021})}\BibitemShut {NoStop}%
\bibitem [{\citenamefont {Errea}\ \emph {et~al.}(2013)\citenamefont {Errea}, \citenamefont {Calandra},\ and\ \citenamefont {Mauri}}]{errea2013first}%
  \BibitemOpen
  \bibfield  {author} {\bibinfo {author} {\bibfnamefont {I.}~\bibnamefont {Errea}}, \bibinfo {author} {\bibfnamefont {M.}~\bibnamefont {Calandra}},\ and\ \bibinfo {author} {\bibfnamefont {F.}~\bibnamefont {Mauri}},\ }\bibfield  {title} {\bibinfo {title} {First-principles theory of anharmonicity and the inverse isotope effect in superconducting palladium-hydride compounds},\ }\href@noop {} {\bibfield  {journal} {\bibinfo  {journal} {Phys. Rev. Lett.}\ }\textbf {\bibinfo {volume} {111}},\ \bibinfo {pages} {177002} (\bibinfo {year} {2013})}\BibitemShut {NoStop}%
\bibitem [{\citenamefont {Zacharias}\ and\ \citenamefont {Giustino}(2020)}]{zacharias2020theory}%
  \BibitemOpen
  \bibfield  {author} {\bibinfo {author} {\bibfnamefont {M.}~\bibnamefont {Zacharias}}\ and\ \bibinfo {author} {\bibfnamefont {F.}~\bibnamefont {Giustino}},\ }\bibfield  {title} {\bibinfo {title} {Theory of the special displacement method for electronic structure calculations at finite temperature},\ }\href@noop {} {\bibfield  {journal} {\bibinfo  {journal} {Phys. Rev. Research}\ }\textbf {\bibinfo {volume} {2}},\ \bibinfo {pages} {013357} (\bibinfo {year} {2020})}\BibitemShut {NoStop}%
\bibitem [{\citenamefont {Zacharias}\ \emph {et~al.}(2023)\citenamefont {Zacharias}, \citenamefont {Volonakis}, \citenamefont {Giustino},\ and\ \citenamefont {Even}}]{zacharias2023anharmonic}%
  \BibitemOpen
  \bibfield  {author} {\bibinfo {author} {\bibfnamefont {M.}~\bibnamefont {Zacharias}}, \bibinfo {author} {\bibfnamefont {G.}~\bibnamefont {Volonakis}}, \bibinfo {author} {\bibfnamefont {F.}~\bibnamefont {Giustino}},\ and\ \bibinfo {author} {\bibfnamefont {J.}~\bibnamefont {Even}},\ }\bibfield  {title} {\bibinfo {title} {Anharmonic lattice dynamics via the special displacement method},\ }\href@noop {} {\bibfield  {journal} {\bibinfo  {journal} {Phys. Rev. B}\ }\textbf {\bibinfo {volume} {108}},\ \bibinfo {pages} {035155} (\bibinfo {year} {2023})}\BibitemShut {NoStop}%
\end{thebibliography}
\end{document}